\documentclass[authoryear,12pt]{elsarticle}
\makeatletter
   \def\ps@pprintTitle{%
      \let\@oddhead\@empty
      \let\@evenhead\@empty
         \def\@oddfoot{\centerline{\thepage}}%
      \let\@evenfoot\@oddfoot
   }
\makeatother
\usepackage[T1]{fontenc}
\usepackage[utf8]{inputenc}

\usepackage{amsmath}
\newcommand\abs[1]{\left|#1\right|}
\usepackage{amssymb}
\usepackage{subfiles} 
\usepackage{subcaption}
\usepackage{multicol}
\usepackage{bm}
\usepackage{graphicx}
\usepackage[export]{adjustbox}
\usepackage[font=footnotesize]{caption}
\usepackage{float} 
\usepackage{placeins}
\usepackage[dvipsnames]{xcolor}

\definecolor{red}{rgb}{1,0,0}
\definecolor{black}{rgb}{0,0,0}
\definecolor{codegreen}{rgb}{0,0.6,0}
\definecolor{codegray}{rgb}{0.5,0.5,0.5}
\definecolor{codepurple}{rgb}{0.65,0,0.82}
\definecolor{backcolour}{rgb}{0.97,0.97,0.97}
\definecolor{codeorange}{rgb}{1,0.31,0}
\usepackage{longtable}

\usepackage{xparse}
        \ExplSyntaxOn
        \NewDocumentCommand{\convertto}{mm}
         {  \texttt{#2~=~\fp_to_decimal:n { (#2)/(1#1) }#1} }
         \ExplSyntaxOff

\def\R{\textsuperscript{\textregistered}}
\def\tcf{\centering\footnotesize}

\usepackage[per-mode=symbol]{siunitx}

\usepackage{amssymb,amsmath,stmaryrd,footmisc}
\usepackage{esint}
\usepackage{placeins}

\usepackage{xcolor}
\usepackage{epstopdf}
\usepackage{natbib}
\usepackage{nohyperref}
\usepackage{graphicx}
 \usepackage{setspace}
\usepackage{color}
\usepackage{verbatim}
 \usepackage[usenames,dvipsnames]{pstricks}
 \usepackage{epsfig}
 \usepackage{pst-grad} 
 \usepackage{pst-plot} 
 \usepackage{fancyref}

\usepackage{etoolbox}
\usepackage{mathtools}
\preto\subequations{\ifhmode\unskip\fi}

\usepackage{physics}
\usepackage{amsmath}

\usepackage{placeins}
\usepackage{siunitx}

\newlength{\tleft}
 \newlength{\tright}
\def\nten#1{\mathbf{#1}}

\def\dD0{\mathcal{\partial D}_0}
\def\D0{\mathcal{D}_0}

\def\dV0{\dd{V_0}}

\def\dA0{\dd{A_0}}
\def\lam{\lambda}

\journal{Journal of the Mechanics and Physics of Solids}
\usepackage{cleveref}

\begin{document}

\begin{frontmatter}

\title{Probing local nonlinear viscoelastic properties in soft materials}

\author[label1]{S. Chockalingam}
\author[label2,label4]{C. Roth}
\author[label2]{T. Henzel}
\author[label2,label3]{T. Cohen\corref{cor1}}
\cortext[cor1]{Corresponding author: talco@mit.edu}

\address[label1]{Massachusetts Institute of Technology, Department of Aeronautics and Astronautics, Cambridge, MA, 02139, USA}
\address[label2]{Massachusetts Institute of Technology, Department of Civil and Environmental Engineering, Cambridge, MA, 02139, USA}
\address[label4]{École Polytechnique Fédérale de Lausanne (EPFL), Institute of Mechanical Engineering, 1015 Lausanne, Switzerland }
\address[label3]{Massachusetts Institute of Technology, Department of Mechanical Engineering, Cambridge, MA, 02139, USA}



\begin{abstract}
Minimally invasive experimental methods that can measure local rate dependent mechanical properties are essential in  understanding the behaviour of soft and biological materials in a wide range of applications. Needle based measurement techniques such as Cavitation Rheology \citep{zimberlin_cavitation_2007} and Volume Controlled Cavity Expansion (VCCE, \cite{vcce-raayai}), allow for  minimally invasive local mechanical testing, but have been limited to measuring the elastic material properties. Here, we propose several enhancements to the VCCE technique to adapt it for characterization of viscoelastic response at low to medium stretch rates ($\num{e-2}$ - $1$ s${}^{-1}$). Through a carefully designed loading protocol, the proposed technique performs several cycles of expansion-relaxation at controlled stretch rates in a cavity expansion setting and then employs a large deformation viscoelastic model to capture the measured material response.  Application of the technique to soft PDMS rubber reveals significant rate dependent material response with high precision and repeatability, while isolating equilibrated states that are used to directly infer the quasistatic elastic modulus. The technique is further established by demonstrating its ability to capture changes in the rate dependent material response of a tuneable PDMS system. The measured viscoelastic properties of soft PDMS samples are used to explain earlier reports of rate insensitive material response by needle based methods:  it is demonstrated that the conventional use of constant volumetric rate cavity expansion can induce high stretch rates  that lead to viscoelastic stiffening and an illusion of rate insensitive material response. We thus conclude with a cautionary note on possible overestimation of the quasistatic elastic modulus in previous studies and suggest that the stretch rate controlled expansion protocol, proposed in this work, is essential for accurate estimation of both quasistatic and dynamic material parameters.
\end{abstract}


\end{frontmatter}


\section{Introduction}
 Mechanical characterisation of soft and biological materials
is important in several applications including tissue engineering \citep{engler2004myotubes,kong2005non,vedadghavami2017manufacturing}, food science \citep{finney1967dynamic,solomon2007modeling}, disease detection \citep{yeh2002elastic,paszek2005tensional,samani2007inverse,last2011elastic} and study of biological processes such as growth and morphogenesis \citep{budday2014role,von2020morphogenesis}. Biological tissues are often heterogeneous and their mechanical properties can change significantly when removed from their native environment \citep{nickerson2008rheological}. Experimental techniques that can measure the mechanical properties of biological materials both locally and \textit{in vivo} are thus essential for accurate characterisation of the material response. For this reason, the Cavitation Rheology technique has emerged as a popular choice for estimation of local elastic material properties of soft and biological materials \citep{zimberlin_cavitation_2007,zimberlin2010cavitation,zimberlin2010water,cui2011cavitation,crosby2011blowing,delbos2012cavity,chin2013cavitation,blumlein2017mechanical,polio2018cross,fuentes2019using}.  In this method, by pressure-controlled inflation, a cavity is expanded in the material at the tip of a needle. Through this inflation process, the pressure inside the cavity reaches a maximum that is assumed to correspond to the theoretically predicted elastic cavitation instability limit, which is used to determine the elastic modulus. Despite its success, the Cavitation Rheology technique often
results in fracture of the sample prior to reaching the cavitation instability limit and relies on an \textit{a priori} assumption on the constitutive response. The Volume Controlled Cavity Expansion (VCCE) technique \citep{vcce-raayai,raayai2019capturing} remedies these problems by performing a volume controlled expansion of the cavity using injection of an incompressible fluid and then using the pressure-volume data (prior to fracture) to extract the nonlinear elastic material properties, as recently demonstrated in application for characterising brain tissue response \citep{vcce_brain}. 
 
While application of needle based techniques has been demonstrated for the measurement of rate independent properties, soft and biological materials often exhibit viscoelastic response, which plays a role in several applications such as disease detection \citep{streitberger2011vivo,tram2018rheological}, study of adhesion behaviour  \citep{castellanos2011effect,prowse2011effects,reza2014effect}, understanding tissue response when subjected to high-intensity focused ultrasound \citep{zilonova2018bubble}, and study of skin pain sensation \citep{liu2015effect}. These applications emphasize the need for reliable experimental techniques that can extract the local nonlinear viscoelastic material properties for soft and biological materials. However, conventional viscoelastic testing methods suffer from several drawbacks when it comes to testing of such materials. For example, uniaxial and simple shear viscoelastic testing \citep{mao2017large,budday2017viscoelastic} of soft materials need to overcome several challenges: sample preparation in specific shapes, boundary effects, and inhomogeneous deformation \citep{rashid2012inhomogeneous,budday2017viscoelastic}. Moreover, for testing of biological tissues, they require the specimen to be cut and taken out of its native environment, negating the possibility of \textit{in vivo} testing. Additionally, only bulk properties of the tissues can be extracted which is not ideal for mechanical characterisation of biological materials that are usually hetereogeneous. Alternatively, small-scale indentation based techniques allow for local viscoelastic material testing \citep{balooch1998viscoelastic,zheng1999extraction,mahaffy2004quantitative,vanlandingham2005viscoelastic,hu2010using,budday2015mechanical}. However, accurate determination of material properties beyond the linear elastic regime  necessitate sophisticated contact mechanics models to be employed which are not always readily available \citep{lin2009spherical,style2013surface}.

Given the limitations of the conventional testing methods described above, an alternative would be to consider experimental methods for local viscoelastic testing in a cavity expansion setting, as suggested by \cite{cohen2015dynamic}. Inertial Microcavitation Rheometry (IMR) \citep{estrada2018high,yang2020extracting} is one such recently developed technique that can be used for local viscoelastic material characterisation at ultra-high stretch rates ($10^3-10^8$ s${}^{-1}$). It uses laser pulses to generate a cavity within the material sample, the ensuing dynamic cavity motion is actively tracked visually and modelled using viscoelastic governing dynamics. The technique is restricted by the requirement of transparent material samples to be able to visually track the cavity motion. Additionally, there is neither an independent measure of the pressure response nor a way to control the stretch rates. Therefore IMR is not a suitable technique for viscoelastic material testing at low to medium stretch rates. Other cavity expansion and instability based techniques such as Cavitation Rheology and VCCE would have to be modified for viscoelastic material testing, as the influence of viscosity can significantly affect both the cavitation instability limit and the pressure response during cavity expansion \citep{cohen2015dynamic, kumar2017some}.

From the discussion of the strengths and weaknesses of the different experimental techniques above, it is apparent that adapting the VCCE technique for rate dependent testing would provide a promising approach for local and \textit{in vivo} characterisation of nonlinear viscoelastic material response at moderate stretch rates. Thus, in this paper, we propose several modifications to the VCCE technique to adapt it for the viscoelastic characterisation of soft solids at low to medium stretch rates ($\num{e-2}$ - $1$ s${}^{-1}$). The modified method performs stretch rate controlled cavity expansion experiments and employs a large deformation nonlinear viscoelastic constitutive model to characterise the experimentally observed material response.

The paper is organized as follows:  we begin with some preliminary definitions and an overview of the proposed experimental method in \Cref{sec:problemsetting}. This is followed by a detailed description of the experimental technique in \Cref{sec:expt_method}. In \Cref{sec:Modelresult}, we present the experimental results for a representative material sample. Then in  \Cref{sec:Governing_eqns}, we set up the governing equations and describe the employed large deformation viscoelastic constitutive model. Subsequently in \Cref{sec:fittedresults}, the procedure for estimation of the material parameters of the constitutive model from experimental results is described using the representative results from \Cref{sec:Modelresult}. The fitted material parameters are used to explain the rate insensitive material response observed in earlier cavity expansion based studies. Finally, in \Cref{sec:Oil_expts}, we demonstrate the potential of the experimental method by studying the  response of PDMS  samples with tuneable viscoelasticity.  We provide some concluding remarks in \Cref{sec:Conclusions}.

\section{Preliminaries and overview of experimental method}
\label{sec:problemsetting}
\begin{figure}
    \centering
    \includegraphics[width=\textwidth]{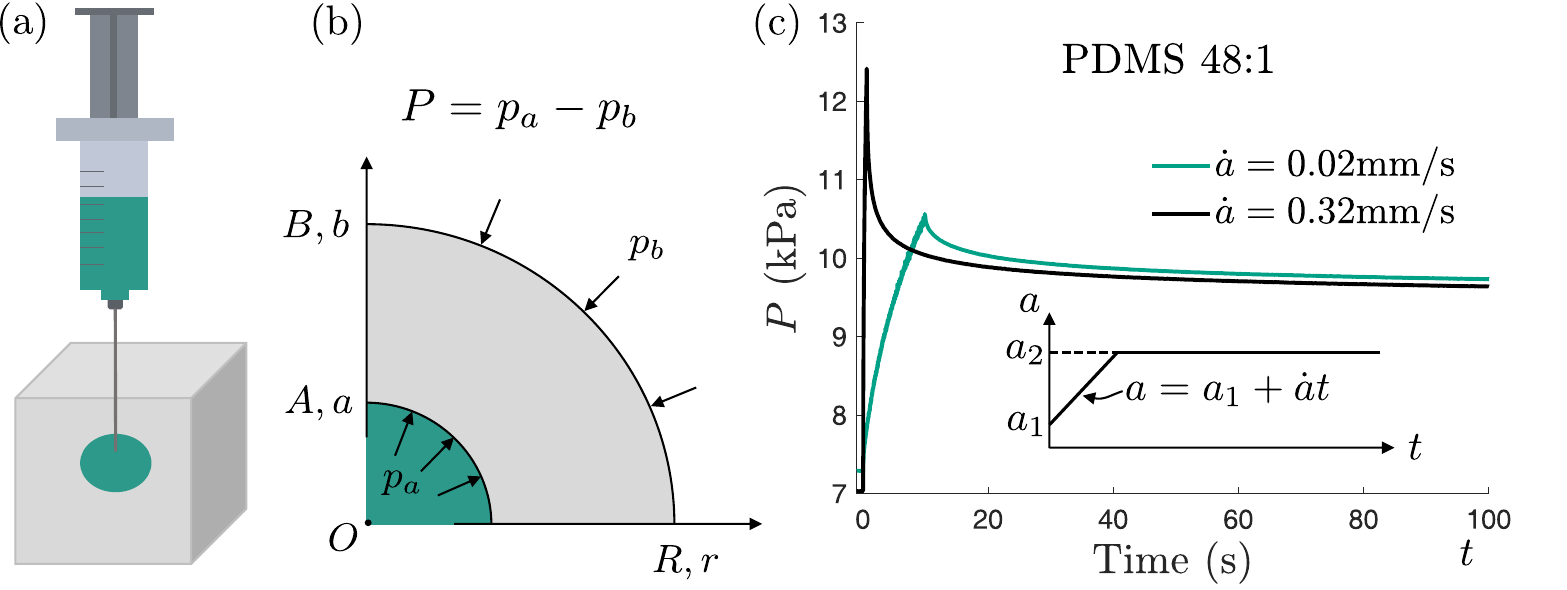}
    \caption{(a) Illustration of Volume Controlled Cavity Expansion inside a soft material sample, through injection of incompressible fluid.  (b) Schematic illustration of the cavity expansion problem. The gauge pressure $P$ is defined as $P = p_a - p_b$ where $p_a$ is the pressure applied on the cavity wall and $p_b = p_{atm}$ is the atmospheric pressure acting on the outer surface of the solid. (c) Rate dependent material response observed in cavity expansion-relaxation experiments for PDMS rubber sample with base:cross-linker ratio of 48:1. The effective cavity size, $a$, is increased from a value of $a_1 = 0.6~\textrm{mm}$ at an initial relaxed state to a final size of $a_2 = 0.8~\textrm{mm}$ at a constant cavity expansion rate ($\dot{a} = $const). Plot of the resulting gauge pressure versus time is shown for two different expansion rates. }
    \label{fig:intro}
\end{figure}
 Uniaxial testing methods commonly probe viscoelasticity using stress relaxation experiments wherein the material sample is dynamically elongated and held at a constant stretch. Then, the stress rapidly increases to a peak value followed by relaxation to an equilibrium value over some characteristic timsecale. Dynamic unloading of the same sample to a constant stretch causes the stress to rapidly drop, followed by a viscoelastic recovery to an equilibrium value. The proposed experimental method in this paper performs analogous stress relaxation/recovery experiments in the spherical expansion setting of a small cavity.

 The experimental method begins with the insertion of a syringe needle into the soft material sample, which creates an initial defect/cavity at the tip of the needle. The cavity is then expanded by performing volume controlled injection of an incompressible and immiscible fluid, as illustrated in \Cref{fig:intro}(a). The pressure inside the cavity is actively measured throughout the experiment. Assuming that the cavity expansion is spherically symmetric, we can define an effective cavity radius, $a$, based on the injected volume of fluid $V$ so that $a = \left({3 V}/{4 \pi} \right)^{1/3}$. The effective stress free size of the initial defect is denoted by $A$ and is retroactively estimated from the pressure-volume data, as later described in \Cref{sec:fittedresults}. 
 An effective circumferential stretch at the cavity wall, $\lam_a$, is then defined as $\lam_a = a/A$.
 
 We define the gauge pressure $P$ as $P = p_a - p_b$, where $p_a$ is the pressure applied on the cavity wall at the interface between the injected fluid and the solid, and $p_b = p_{atm}$ is the atmospheric pressure acting on the outer surface of the solid. The experimental method prescribes a loading protocol (time profile of specified effective cavity size $a$) that involves several cycles of cavity expansion-relaxation and retraction-recovery between two fixed effective cavity sizes. The resulting experimental gauge pressure profiles are captured using a large deformation nonlinear viscoelastic constitutive model to extract the local viscoelastic material properties.
 
 To motivate the proposed experimental method, we present a precursory experimental result that demonstrates significant viscoelastic effects in the material response of soft PDMS rubber (base:cross-linker ratio of 48:1). The prescribed loading expands an initially relaxed cavity with effective size $a_1=0.6$ mm, to a final size $a_2=0.8$ mm, at a fixed cavity expansion rate ($\dot{a} =$ const.), and then holds the cavity size at $a_2$ while allowing the material to relax. The resulting experimentally measured gauge pressure is plotted as function of time, for two different cavity expansion rates, in \Cref{fig:intro}(c). The gauge pressure monotonically increases during expansion to a peak pressure and then relaxes to an equilibrium value when the effective cavity size is held constant. The rate dependence of the material response is apparent from the larger peak pressure for the faster cavity expansion rate and the significant viscoelastic relaxation. The equilibrated pressure which is approximately same for both cavity expansion rates, characterises the equilibrium elastic response of the material.

The successful generation of such experimental pressure curves with accuracy, precision, and repeatability, involves several components including isolation of a fracture-free stretch range, elimination of Mullins effect, and accounting for surface tension effects and dynamic pressure losses. Also important is the identification of a loading protocol that facilitates observation of rate dependent material response while isolating consistent/repeatable equilibrated pressures. These equilibrated pressures are used to extract quasistatic properties of the material whereas the pressure-time curves are used to estimate the dynamic material properties. In the following section, the experimental method is described in detail, including its various individual components that account for the different factors discussed above.

\section{Experimental method}
\label{sec:expt_method}
 In the following subsections we describe different aspects of the full experimental protocol which is finally summarised in \Cref{tab_fullp}.

\subsection{Setup}
\begin{figure}
\centering
\hspace*{-3.5cm}\includegraphics[height=9cm]{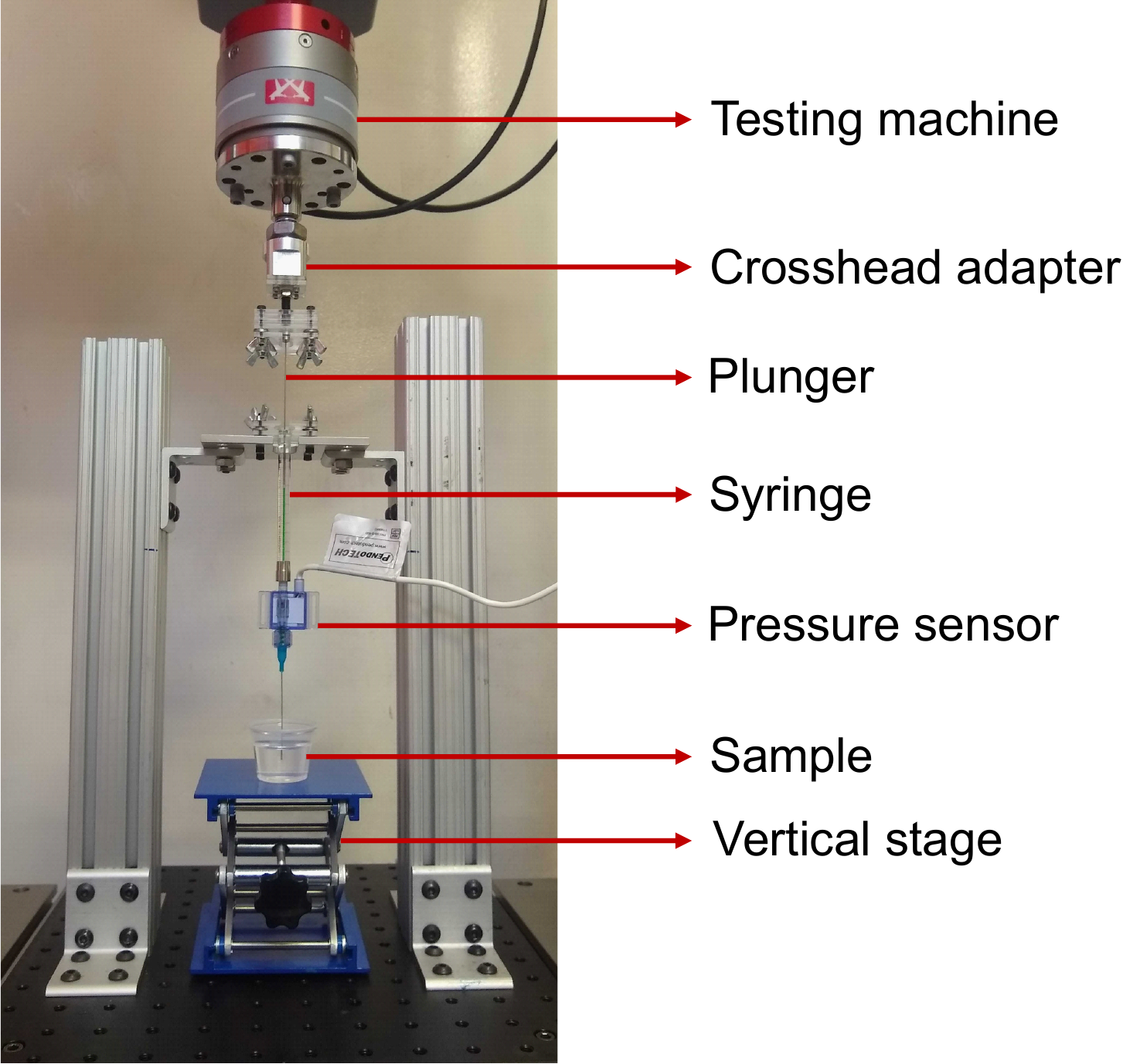}
\caption{Experimental setup}
\label{fig_pictsetup}
\end{figure}

 The experimental setup is shown in \Cref{fig_pictsetup}. It consists of a stationary syringe barrel held by fixed supports and a movable plunger actuated by an Instron$\R$ Dynamic Test Instrument. The syringe is filled with an incompressible fluid (water in this paper) and the plunger displacements control the volume of fluid ejected. An in-line pressure sensor connects to the tip of the syringe and hosts the needle on the other outlet with standard luer connections. The sensor is calibrated to the testing machine as an external transducer. The material sample being studied is placed on a vertical stage and raised until the needle penetrates the material surface. We then perform volume controlled fluid injection and study the dynamic pressure response inside the fluid filled cavity.

A key difference from the VCCE experimental setup in \cite{vcce-raayai} is the measurement of pressure through a pressure sensor rather than through the calibrated forces\footnote{Calibration required elimination of frictional forces.} measured by the testing machine, which results in more reliable and precise pressure measurements. This change also allows the use of a gas-tight high accuracy $50~\mu \textrm{L}$ syringe, compared to the $3~\textrm{mL}$ syringe in \cite{vcce-raayai}. The resulting near hundred times reduction in cross sectional area leads to significantly more precise volume control. Use of smaller cross section syringes along with force measurements through the mechanical testing machine is prohibited by unreliable calibration of much higher frictional forces. Hardware components used in the experiment are listed in \Cref{tab_hardware}.

\begin{table}
\tcf
\caption[Technical Hardware Components]{Hardware components including precision where applicable.}  \label{tab_hardware}
\begin{tabular}{p{0.16\textwidth} p{0.7\textwidth}}
\hline
Testing machine & Dynamic Instron\R ElectroPlus\textsuperscript{TM} E3000\newline
Guaranteed displacement precision: $\pm0.02$ mm \\
Syringe & Hamilton\R Gastight Syringe Model 1705 PTFE Luer (TLL),\newline
Capacity $50~\mu \textrm{L}$, surface area $A_{s}=0.83~\si{\square\mm}$,\newline Max pressure 6.9 MPa = 1000 psi \\
Needle & Stainless-Steel  Dispensing Needles, \newline Blunt Tip Gauge 25 $\varnothing _{out }= 0.51~\si{mm}, \varnothing _{in }= 0.30~\si{mm}$\\ 
Pressure Sensor & Pendotech\R PRESS-S-000, Range: 79 \si{kPa} - 520 \si{kPa},\newline Accuracy: $\mathrm{\pm2\%~\forall p<40~  \si{kPa}}$,\newline Response time $1~$ms\\
\hline
\end{tabular}
\end{table}

\subsection{Effective cavity size}
\label{subsec:cavitysize}
The effective size of the cavity at any time can be directly calculated from the volume of fluid injected into the material,
\begin{gather}
a=\sqrt[]{\frac{3}{4\pi}{A_{s} z}}
\end{gather}
 where $z$ is the displacement of the syringe from its initial position. A smaller syringe cross section area $A_{s}$ leads to higher accuracy of the measure of $a$ for given displacement precision of the mechanical testing machine.

\subsection{Pressure correction}
\label{subsec:pressure_correct}
The raw pressure data measured by the sensor, $p_m$, is first corrected by the ambient air pressure $ p_{atm}$, to obtain the measured gauge pressure $P_m$,
\begin{equation}
    P_m = p_m - p_{atm}
\end{equation}
\begin{figure}
    \centering
    \includegraphics[width=\textwidth]{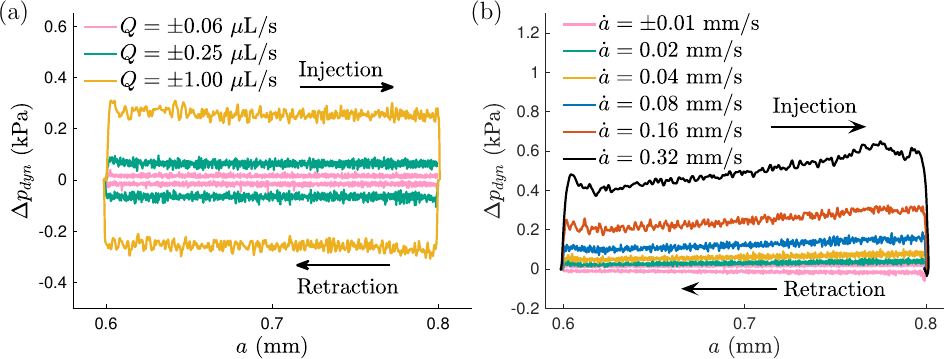}
    \caption{Example experimental calibration curves for dynamic pressure drop correction.}
    \label{fig:my_calibration}
\end{figure}
Although we are interested in measuring the pressure applied on the cavity wall, the pressure is being measured at some distance from the tip of the needle (\Cref{fig_pictsetup}). While the pressure difference caused due to the weight of the fluid column is negligible for the scale of pressures in our experiments, the dynamic pressure loss caused by the fluid flow in the needle should be accounted for, as well as the possible effect of surface tension. To account for the dynamic pressure loss, we perform calibration experiments by injecting and retracting working fluid from a container at all the different combinations of cavity sizes and expansion rates used in an experiment. Plots of the dynamic drop in pressure ($\Delta p_{dyn}$) from sample calibration experiments are shown in \Cref{fig:my_calibration}, for the cases of constant fluid volume flow rate $Q$ and constant cavity expansion rate $\dot{a}$. It can be seen that the dynamic pressure drop is constant for given flow rate $Q$ and higher for higher $Q$. The drop is non-uniform and increases with $a$ and $\dot{a}$ for constant cavity expansion rates (since $Q = 4 \pi a^2 \dot{a}$). The gauge pressure inside the fluid in the cavity, $P_f$, is then estimated from the measured gauge pressure $P_m$ as 
\begin{equation}
\label{eq:corrected_P1}
    P_f(t) = P_m(t) - \Delta p_{dyn}(a(t),\dot{a}(t)) 
\end{equation}
Calibration for the dynamic pressure correction is performed before every single test. Retraction at even moderate rates can cause detachment between the plunger and the fluid, and thus \textit{we restrict retraction rates to  $\dot{a} > -0.01~\si{mm/s}$}. Additionally, at the rates considered here, pressure difference generated due to inertial effects in the fluid container is insignificant. 
Finally, the gauge pressure at the cavity wall, $P$, can be estimated from the gauge pressure inside the fluid in the cavity, $P_f$, by accounting for the effect of surface tension as \citep{mishra2018effect}
\begin{equation}
\label{eq:corrected_P}
    P(t) = P_f(t) - \frac{2 \gamma}{a(t)}
\end{equation}
where $\gamma$ is the surface energy for the fluid-solid interface.  Henceforth, pressure refers to the cavity wall gauge pressure $P$, in \cref{eq:corrected_P}.

\subsection{Experiment initiation procedure}

To ensure repeatability, in this work we employ a strict protocol to initiate the cavity expansion experiments from a well defined relaxed state. First, the pressure at the tip of the needle is actively monitored as the needle penetrates into the solid  by raising the vertical stage (\Cref{fig_pictsetup}). As the solid surface deforms, there is a monotonic increase in pressure, followed by a sudden drop when the surface ruptures. The needle is then slightly retracted, by lowering the stage, to relax most of the compression
below the needle ($p\gtrapprox0$) and to ensure that the relaxing material does not block the needle entry. Since the measurement accuracy for small cavity size is unsatisfactory, the cavity is inflated to a size of $a=0.5~\si{mm}$ at a very slow expansion rate of $\dot{a} < 0.01~\si{mm/s}$.  In this state, the pressure is given time to relax (1 - 4 hours). For the materials investigated here, cavities formed this way have repeatably proven to be of visually spherical shape and to conserve the spherical geometry during expansion. Hence, this serves as a repeatable well-defined initial condition for all experiments.

\subsection{Isolating a fracture free stretch range}
\begin{figure}
    \centering
    \includegraphics[width=\textwidth]{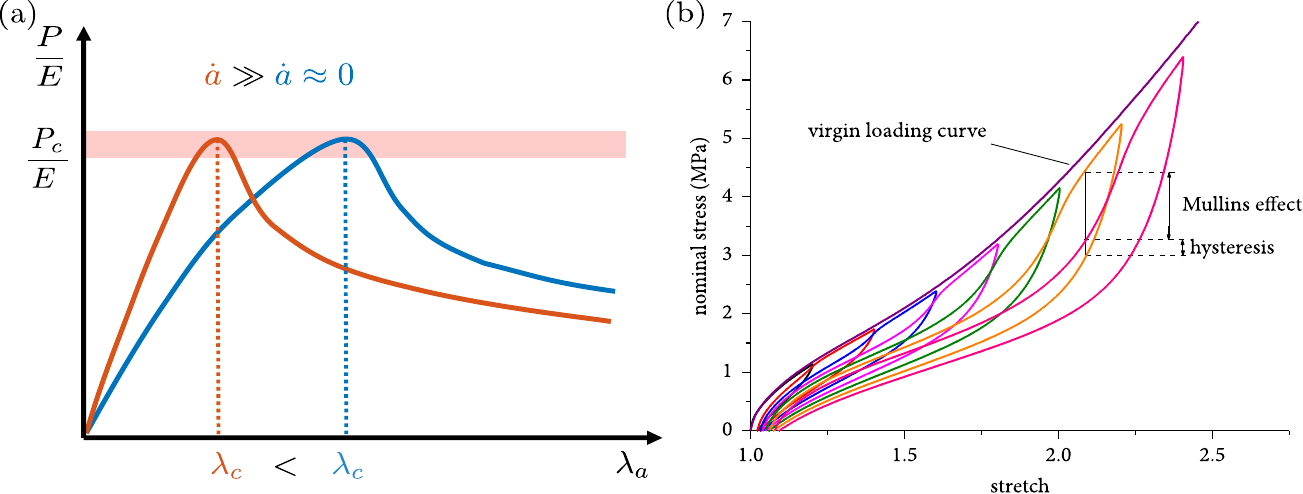}
    \caption{(a) Qualitative illustration of pressure versus cavity stretch  before and after onset of fracture, at $\lambda_c$. Fracture, identified by a sharp drop in pressure, has been observed to occur at almost the same pressure $P_{c}$ for the expansion rates considered here. Increased elastic material resistance at higher expansion rates results in lower values of  $\lambda_{c}$. Thus the size of the fracture free cavity stretch range, $\lambda_a \in [1 \ \lambda_{c})$, reduces at higher expansion rates. (b) Representative Mullins effect observed in cyclic uniaxial testing of  natural/styrene-butadiene (NSBR) rubber \citep{huang2019pseudo}. The different colours correspond to different loading cycles. }
    \label{fig_elastic}
\end{figure}

An important component of the proposed method is the isolation of a cavity stretch range in which the material does not fracture. As discussed in  \cite{intimate_2019}, fracture and elastic cavity expansion are intimately coupled. At first when the cavity expands, elasticity dominates the pressure response until a critical pressure $P_{c}$ is reached. After this point, the pressure sharply drops as fracture and elastic resistance simultaneously influence the pressure response, as illustrated in \Cref{fig_elastic}(a). In this work, we define the cavity stretch range prior to the first sharp pressure drop as fracture free. For the material systems studied here, we observe that  $P_{c}$ appears to be roughly independent of expansion rate $\dot{a}$. A viscoelastic material demonstrates dynamic amplification of elastic resistance, i.e the pressure versus cavity stretch curve steepens compared to the quasistatic limit, as shown in \Cref{fig_elastic}(a). This implies that the fracture free cavity stretch range is smaller for higher expansion rates. We experimentally determine the critical effective cavity radius  $a_{c}$ that initiates fracture at the highest cavity expansion rate we want to probe at. Assuming the effective stress free size of the initial defect, $A$, is of similar order between probings, $a_{c}$  defines an upper limit of cavity size to be probed to remain fracture free.

\subsection{Mullins effect}

The Mullins effect  is often used to describe stress softening after large deformations in soft rubbery materials \citep{clement_mullins_2001,mullin,diani_review_2009}. Typically studied for cyclic uniaxial tests (\Cref{fig_elastic}(b)), the effect is characterised by a softening behaviour that appears when a stretch value that exceeds values in previous cycles is accessed.  Within a previously accessed stretch range, the material responses coincide during the following cycles, aside from a fatigue effect. When the stretch arrives at the maximum value previously applied, the material response reunites with the first uniaxial tension test path.

\begin{figure}
    \centering
    \includegraphics[width=\textwidth]{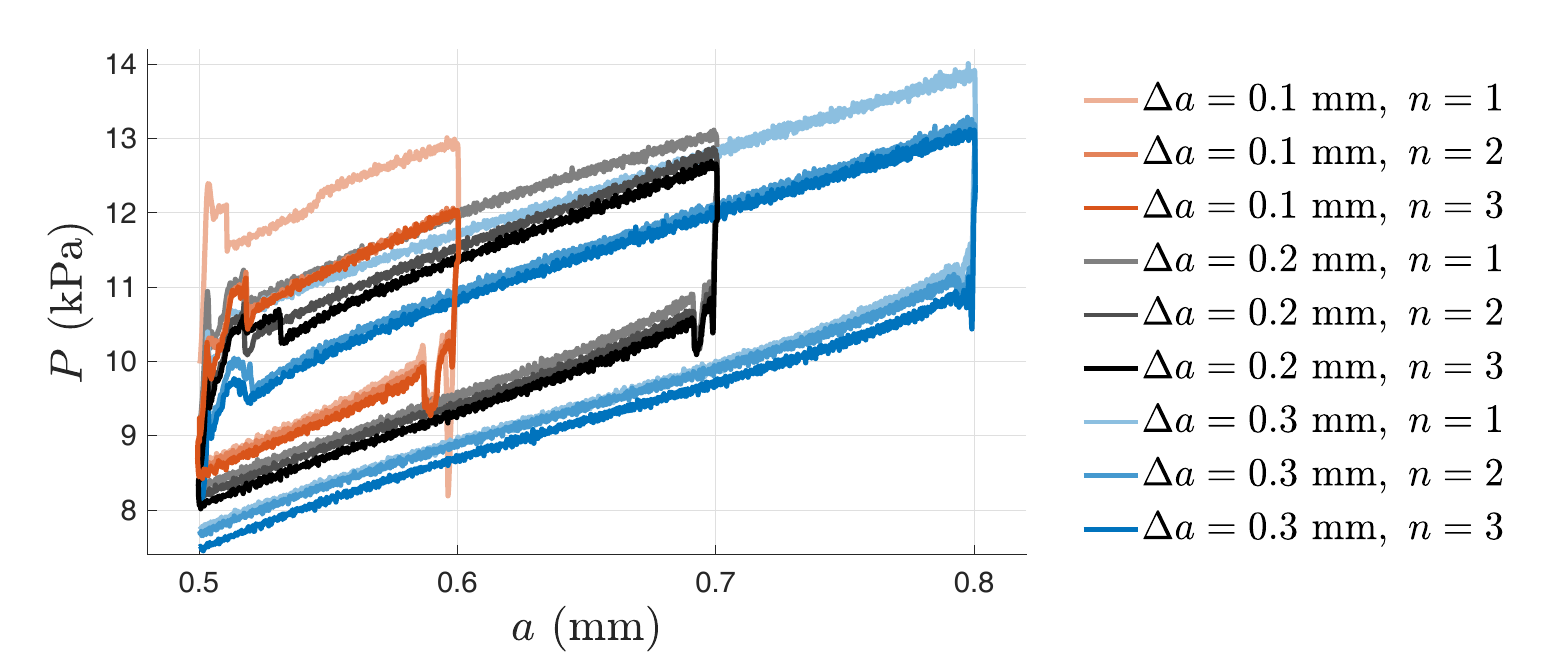}
    \caption{Demonstration of the Mullins effect in cavity expansion setting for PDMS sample S50-00 (see \Cref{sec:Oil_expts}). Within an activated stretch range, the material response changes significantly from the first cycle to the second but thereafter remains unchanged over subsequent cycles. Hysteresis is observed between loading and unloading. The significant change in pressure whenever the loading rate is reversed is an indicator of rate dependent material response.}
    \label{fig:mullins_cavity}
\end{figure}

We demonstrate the presence of this effect in the cavity expansion setting as well. 
Starting at a fully relaxed cavity at $a=0.5~\si{\mm}$, fluid is injected at an expansion rate of $\dot{a}=0.01~\si{mm/s}$ up to $a=0.5+\Delta a\ \si{mm}$.  We then immediately retract at a rate of $\dot{a}=-0.01~\si{mm/s}$ back to $a=0.5~\si{mm}$. Three such cycles are performed for increasing values of $\Delta a$ and before every increase in $\Delta a$, the cavity is given time to relax fully. The resulting pressure profiles for a PDMS sample S50-00, whose composition is defined later in \Cref{sec:Oil_expts}, are shown in \Cref{fig:mullins_cavity}. The parallels with the Mullins effect for uniaxial cyclic testing in \Cref{fig_elastic} can be clearly seen. For any cycle, a large hysteresis is seen between the expansion and retraction curves. The expansion pressure profile demonstrates softening on the second cycle for any given $\Delta a$ but by the third cycle the material responses coincide. The retraction profiles do not demonstrate softening and remain approximately unchanged. In the first cycle ($n=1$) every time $\Delta a$ is increased, the pressure profile follows the profile from the previous cycle. {\textit{Note the significant change in the pressure whenever {${\dot{a}}$} changes sign between expansion and retraction. This is already a clear indicator of viscoelastic rate dependent material response even at this low expansion rate.}} 

While there is no consensus on the physical source or on the mechanical modeling of Mullins effect \citep{diani_review_2009},  the present experimental procedure can be applied to obtain further insight into its manifestation in additional stress states (beyond uniaxial tests). However, this is beyond the scope of the present work. Hence, in our experimental protocol, we eliminate the Mullins effect by pre-loading the material in the required stretch range of any experiment and letting the cavity relax at the maximum stretch. To confirm the removal of Mullins effect, we pre-cycle three times in the required experimental range of $a$ at a very low rate and verify that the hysteresis during expansion is fully removed. 

\subsection{Loading protocol}
\label{sec_final_protocol}

\begin{figure}[h!]
\centering
\includegraphics[width=0.95\textwidth]{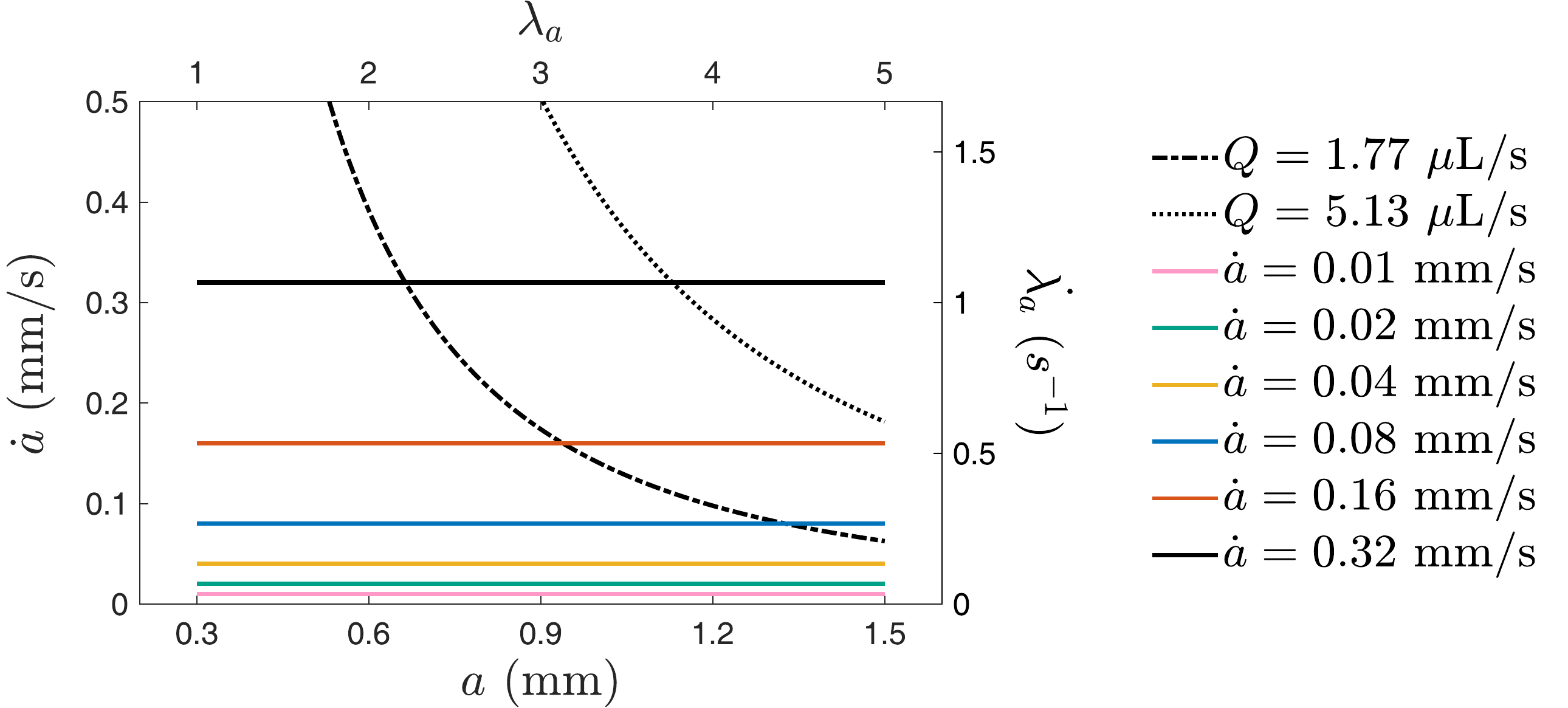}
\caption[Effective Expansion Rates for Constant $Q$ and $\dot{a}$]{Comparison of cavity  expansion rates ($\dot{a}$) and cavity stretch rates ($\dot{\lam}_a = \dot{a}/A$) employed in this paper with those in the experiments of \cite{vcce-raayai}, as a function of effective cavity size. The cavity stretch and stretch rates have been plotted for a nominal value of $A=0.3$ mm. Constant cavity expansion rates ($\dot{a}=\textrm{const}$) result in constant cavity stretch rates ($\dot{\lam}_a =\textrm{const}$). Employing constant volume expansion rate $Q$, as in \cite{vcce-raayai}, results in large variations of cavity stretch rate during expansion and in high stretch rates at small cavity sizes that can lead to saturated material response ill-suited for viscoelastic material characterisation (see \Cref{sec:constQ}). 
}\label{fig_rates}
\end{figure}

A crucial difference between the experimental method in this paper, and existing cavity based experimental approaches, is the use of constant cavity expansion rates ($\dot{a} =\textrm{const}$) instead of the typical constant fluid volume flow rate $Q$. The use of constant volume flow rate $Q\ (= 4\pi a^2 \dot{a})$ in the expansion of a small initial cavity results in  significant variation in the cavity stretch rate $\dot{\lam}_a (= \dot{a}/A)$ during expansion, and in especially high cavity stretch rates at small $a$. This is 
shown in \Cref{fig_rates} for the flow rates employed in \cite{vcce-raayai}. We later demonstrate in \Cref{sec:constQ} that, for the materials used in this work, such volume flow rates (which are typically used in `quasistatic' cavity based experiments) activate high stretch rate response. Hence, the inference that the experiments were conducted quasi-statically is incorrect. Nonetheless, it is confirmed that the pressure profiles show little rate dependence, since the material response saturates for the high stretch rates activated.

After cavity initiation and removal of Mullins effect, as described earlier, we begin our main loading protocol. We cycle at constant cavity expansion rates between effective cavity sizes $a_1 = 0.6~\si{mm}$ and $a_2 = 0.8~\si{mm}$. These cavity sizes comfortably fall within the fracture free range we identified for the material systems considered here and can be modified for other materials. The retraction from  $a_2$ to $a_1$ is always performed at the minimum rate $\dot{a}=-0.01~\si{\mm\per\s}$ to prevent detachment of fluid and plunger whereas the expansion rate is doubled between every expansion from a value of $\dot{a}=0.02~\si{\mm\per\s}$ to $\dot{a}=0.32~\si{\mm\per\s}$. At the end of each expansion and retraction, the cavity is given time of $t_{wait} = 400 ~\si{s}$ to relax. This is clarified in \Cref{tab_fullp} where the entire protocol discussed in this section has been summarized.  

\begin{figure}[h!]
    \centering
    \includegraphics[width=\textwidth]{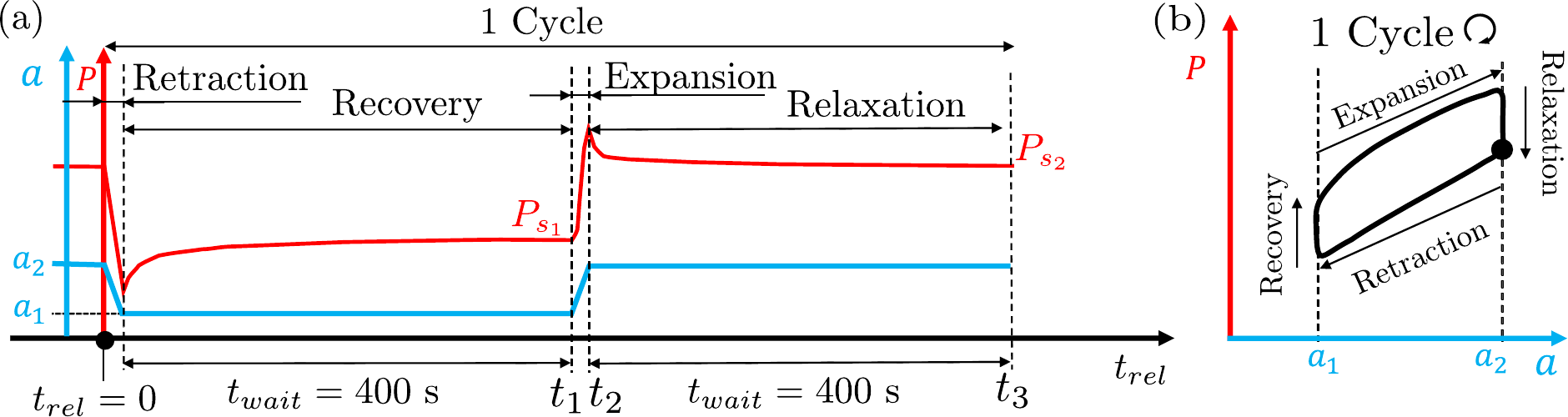}
    \caption{Representative figure of a typical pressure response for one loading cycle in the loading protocol described in \Cref{tab_fullp}. The expansion starts at $t_{rel}=t_1$ and ends at $t_{rel}=t_2$, after which the cavity is allowed to relax upto $t_{rel}=t_3$ before the next loading cycle starts.}
    \label{fig:sample_protocol}
\end{figure}

Representative figures of typical pressure profiles that result from this experimental protocol are shown in \Cref{fig:sample_protocol} for one loading cycle. The time elapsed in a given cycle is identified by $t_{rel}$ which is zeroed at the start of every cycle. The expansion starts at $t_{rel} = t_1$ and ends at $t_{rel} = t_2$ after which the material is allowed to relax at a constant stretch, up to $t_{rel} = t_3$.

\footnotesize{  \begin{longtable}{p{0.23\textwidth}| p{0.7\textwidth}}
\caption{Full experimental protocol for probing viscoelastic rate dependent material response using VCCE.}\label{tab_fullp}\\
                                            
\hline\hline
Assembly: & Fill and install the syringe-sensor assembly while avoiding air entrapments.\\
Calibration for dynamic pressure correction:& Inject needle in container of working fluid and perform full retraction, expansion cycles for all testing cycles. Record $\Delta p_{dyn}(a,\dot{a})$ for pressure correction. \\
Balance:& Remove fluid container, eject fluid droplet to ensure needle is completely filled. \newline Wait $10~\si{\s}$ to determine ambient pressure $p_{atm}$.
\newline Balance Input displacement ($z=0$). \\
Insertion and initiation:& Insert needle in sample till surface ruptures, and immediately retract until $p\gtrapprox0$.
\newline Expand initial cavity to $a=0.5~\si{\mm}$, give time to relax fully ($t_{wait}\approx 1$ hour). \\
Mullins effect removal:& Increase volume to $a_1=0.6~\si{mm}$, let relax fully ($t_{wait}\approx 1$ hour).
\newline Increase volume to $a_2=0.8~\si{mm}$, let relax fully ($t_{wait}\approx 1$ hour).\\
 Cycle 0 (confirmation of Mullins effect removal):& Retract at $\dot{a}=-0.01~\si{\mm\per\s}$ ($a_2 \rightarrow a_1$), $t_{wait}=400~\si{\s}$.
\newline Expand at $\dot{a}=0.01~\si{\mm\per\s}$ ($a_1\rightarrow a_2$),  $t_{wait}=400~\si{\s}$.
\newline
 Repeat 3 times.\\
 Cycles 1 - 5:& For i = 1 to 5, $\dot{a}_1 = 0.02~\si{mm/s}$, do
 \newline \{ Retract at $\dot{a} = -0.01~\si{mm/s}$ ($a_2\rightarrow a_1$),  $t_{wait}=400~\si{\s}$.
\newline \hspace*{0.17cm} Expand at $\dot{a}_i$ ($a_1\rightarrow a_2$), $t_{wait}=400~\si{\s}$.
 \newline \hspace*{0.25cm}$ \dot{a}_{i+1}= 2 \dot{a_i}$ \}\\
\hline\hline
\end{longtable}}                       
\normalsize
\section{Representative experimental results} 
\label{sec:Modelresult}

\begin{figure}[!htb]
    \centering
    \includegraphics[width = \textwidth]{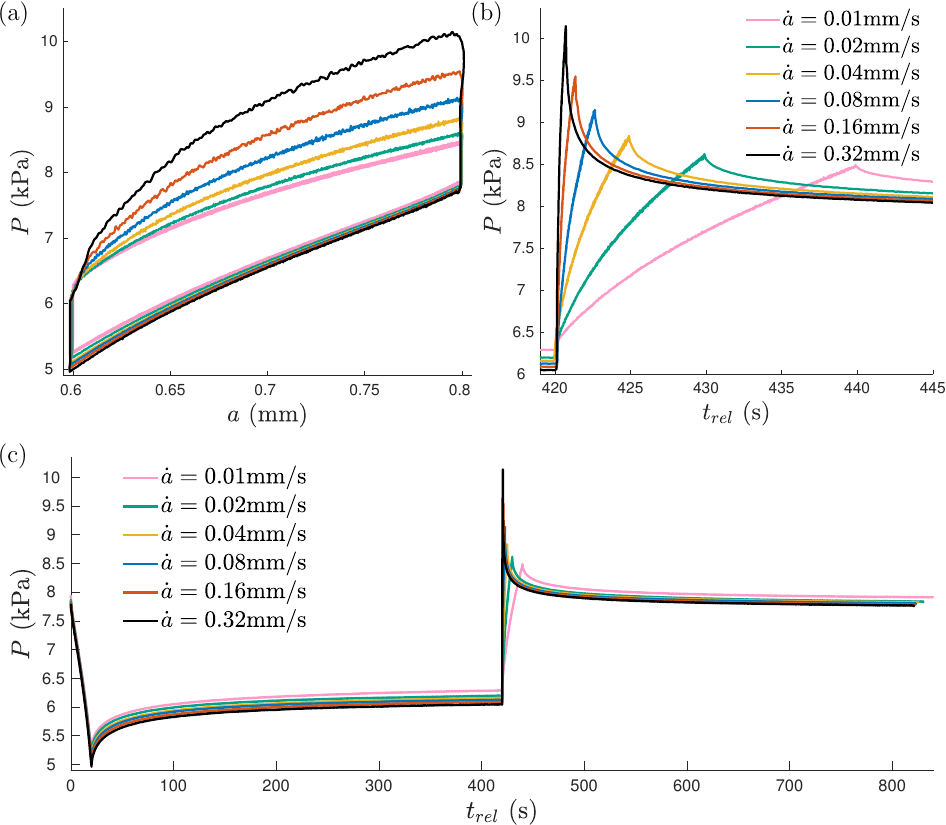}
    \caption{Experimental gauge pressure data for soft PDMS rubber sample with base:cross-linking agent ratio of 50:1 (sample S50-00 of \Cref{sec:Oil_expts}). (a) Pressure data as a function of effective cavity size $a$. (b) Pressure data as a function of time, for the expansion response. (c) Complete pressure-time profiles.  }
    \label{fig:model_result}
\end{figure}

Having defined the full experimental protocol, we first show representative results for soft PDMS (Polydimethylsiloxane) rubber sample (Sylgard 184, Dow Corning) with  base:cross-linking agent ratio of 50:1 (more details on the sample and its  preparation follow in \Cref{sec:Oil_expts}). The surface tension correction is done using $\gamma = 40~\si{mN/m}$ \citep{ismail2009interfacial,fox1947polyorganosiloxanes}. The corrected gauge pressure is plotted in \Cref{fig:model_result}, both as a function of the effective cavity size $a$ and as a function of time. The first thing to note is that pressure profiles for the three cycles of loadings in \Cref{fig:model_result}(a) for the expansion rate $\dot{a}=0.01 \si{mm/s}$ (Cycle 0) coincide. This both confirms the removal of Mullins effect and demonstrates the remarkable precision and control that follows from the experimental protocol detailed in the previous section. \textit{{Furthermore, the change in expansion pressure profile with increasing expansion rate is a clear signature of rate dependent material response}}.
The peak pressure for every cycle, $P_{max},$ is attained whenever the maximum cavity size $a=a_2$ is first reached ($t_{rel}=t_2$) and is higher for higher expansion rates. Since the retraction is always performed at $\dot{a} = -0.01 ~\si{mm/s}$, the retraction pressure profiles nearly coincide. Stress relaxation happens while the cavity size is held constant at $a=a_2$, which causes the pressure to drop with time to an equilibrium value $P_{s_2}$ (also shown in \Cref{fig:sample_protocol}). Conversely, stress recovery occurs while the cavity size is held constant at $a=a_1$, which causes the pressure to increase to an equilibrium value $P_{s_1}$. The relaxation and recovery consistently result in approximately similar equilibrated pressure values over all the cycles, indicating that the equilibrium elastic part of the material response is well captured. There is a however slight decay in the equilibrated pressure values between cycles (<$3\%$ over all cycles). The monotonous nature of the decay seems to indicate there might be fatigue due to damage accumulation over multiple cycles. However, it could also indicate ongoing relaxation at material timescales much larger than the experimental timescale.

We quantitatively qualify the change in material resistance with loading rates by defining the dynamic amplification ratio $R_P$,
\begin{equation}
    R_P = (P_{max}-P_{s_1})/(P_{s_2}-P_{s_1}) \label{eq:DAR}
\end{equation}
For the experimental data shown in \Cref{fig:model_result}, we obtain approximate $R_P$ values of 1.4, 1.5, 1.6, 1.8, 2.0 and 2.4, corresponding to Cycle 0 to Cycle 5. Since $R_P=1$ for quasistatic loading, even the slowest expansion rate considered here shows dynamic amplification of $40 \%$ in the pressure response. Thus, any consideration of the experiment to have been conducted quasistatically would be untenable. Modelling of the rate dependent material response then becomes essential and is the subject of our next section.

\section{Governing equations and generalized nonlinear viscoelastic model}
\label{sec:Governing_eqns}

\subsection{Constitutive response}
\label{subsec:constit}
\begin{figure}
    \centering
    \includegraphics[scale=0.2]{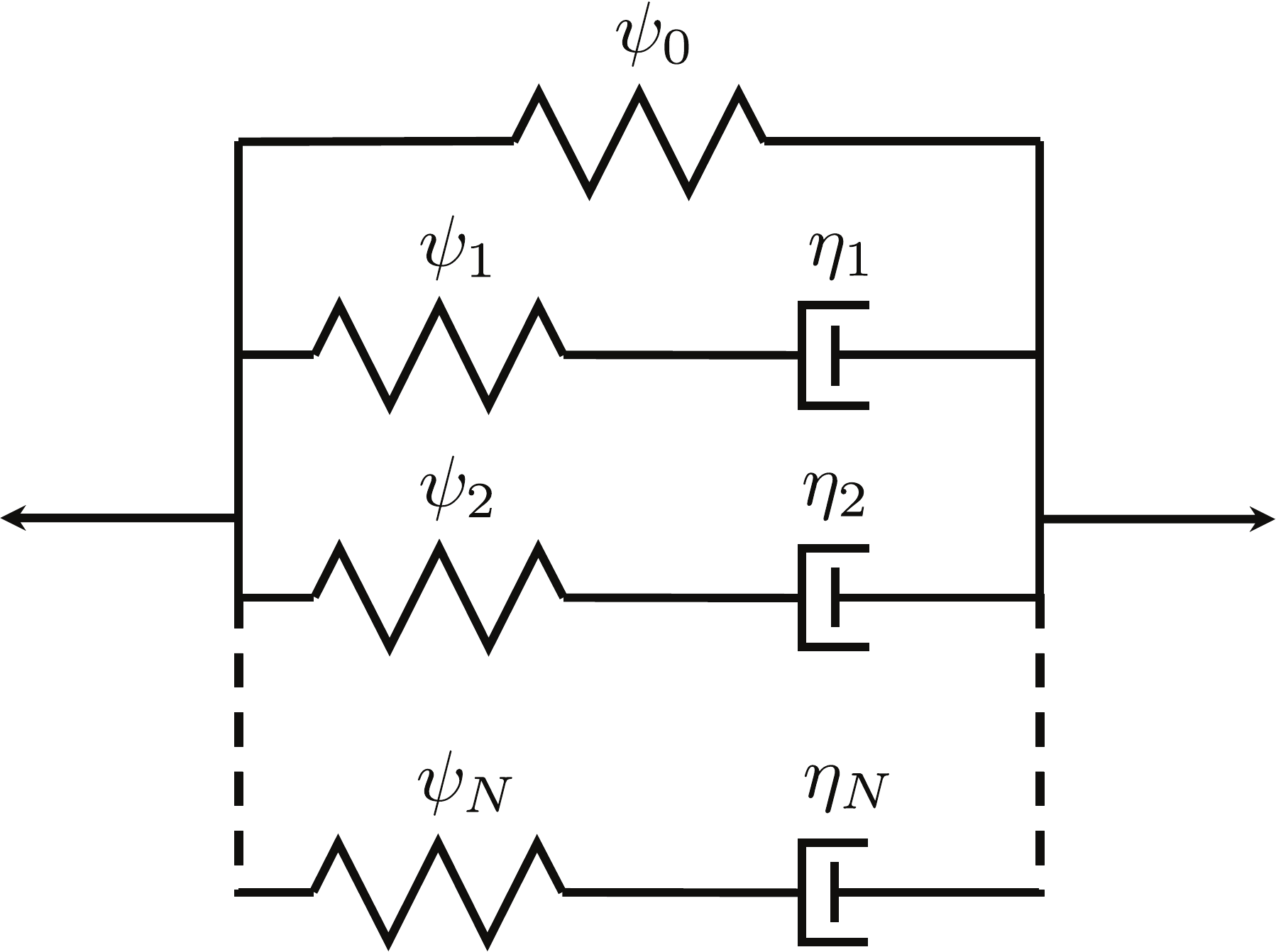}
    \caption{Generalized Maxwell rheological model considered in this paper with an equilibrium branch and  $N$ non-equilibrium branches.  The equilibrium branch is described by a hyperelastic spring with free energy density $\psi_0$ and the $n^{\textrm{th}}$ non-equilibrium branch is described by a hyperelastic spring with free energy density $\psi_n$ and a viscous dashpot with viscosity material parameter $\eta_n$. 
    }
    \label{fig:rheo_model}
\end{figure}
 
To model the viscoelastic response of a general nonlinear isotropic, and incompressible material, we follow the constitutive modelling approach in \cite{kumar2016two,kumar2017some,ghosh2020two} with some modifications.
Motivated by our experimental observations, we consider a generalized Maxwell rheological model with an equilibrium branch and $N$ non-equilibrium branches as shown in \Cref{fig:rheo_model}. To describe the response of each non-equilibrium branch, denoted by $n=1,2,...,N$, the deformation gradient $\nten{F}$ is multiplicatively decomposed as\footnote{Repeated indices do not imply summation throughout this paper.}
\begin{equation}
    \nten{F} = \nten{F}_n^e \nten{F}^v_n \qquad \text{ for }  n=1,2,...,N\label{eq_F_decomp}
\end{equation}
where $\nten{F}^v_n$ is the viscous distortion and $\nten{F}_n^e$ is the non-equilibrium elastic distortion, in the $n^{\textrm{th}}$ non-equilibrium branch. The viscous distortion $\nten{F}^v_n$ is known at $t=0$ and evolves over time by a prescribed kinetic law. All the spring elements in the generalized Maxwell rheological model (\Cref{fig:rheo_model}) are assumed to be incompressible and thus $\textrm{det}(\nten{F})=\textrm{det}(\nten{F}_n^e) =1$ for $n=1,2,...,N$.  The free energy density of the system is assumed to be of the form 
\begin{equation}
\label{free_en_total}
    \psi = \hat{\psi}_0(\nten{F}) + \sum_{n=1}^{N} \hat{\psi}_n(\nten{F}_n^e) = \hat{\psi}_0(\nten{F}) + \sum_{n=1}^{N} \hat{\psi}_n(\nten{F}{\nten{F}^v_n}^{-1}) \
\end{equation}
where $\psi_0$ and $\psi_n$ are the free energy densities of the equilibrium and non-equilibrium branches respectively, and according to \eqref{eq_F_decomp}, $\nten{F}_n^e = \nten{F}{\nten{F}^v_n}^{-1}$. The Cauchy stress tensor $\bm{\sigma}$ is then given as
\begin{equation}\label{cauchy_general_constit}
    \bm{\sigma} = \pdv{\psi}{\nten{F}} \nten{F}^T - p \nten{I}
\end{equation}
where $p$ is an arbitrary hydrostatic pressure associated with the incompressibility constraint. We prescribe the time evolution of the viscous distortion through a kinetic law for $\nten{C}_n^v= {\nten{F}_n^v}^T \nten{F}_n^v$, that satisfies the second law of thermodynamics, isotropic symmetry and incompressiblity requirements, and material frame indifference (see \ref{app:constit100}),
\begin{equation}\label{general_evoln_viscous}
   \nten{\dot{C}}_n^v =  \frac{2}{\eta_n I_{1n}^v}\left( {\nten{{F}}^v_n}^{T} \nten{M}_n \nten{{F}}^v_n - \frac{1}{3}\textrm{tr}(\nten{M}_n)\nten{C}_n^v\right) \quad, \quad \nten{M}_n = {\nten{F}_n^e}^{T}\pdv{\hat{\psi}_n}{\nten{F}^e_n}
\end{equation}
where $I_{1n}^v = \textrm{tr}(\nten{C}^v_n)$, and $\eta_n$ is a positive viscosity material parameter associated with the $n^{th}$ non-equilibrium branch.

\subsection{Spherically symmetric deformation}
\label{subsec:governing_eqn}

The geometry of a finite body with a cavity undergoing spherically symmetric expansion is illustrated in \Cref{fig:intro}(b). The body is loaded by internal and external pressures $p_a$ and $p_b$ respectively, and the gauge pressure $P$ is defined as $P = p_a - p_b$. The radial coordinate in the initial and deformed configurations is defined by $R$ and $r$, respectively. The corresponding inner and outer radii of the body in the undeformed and deformed configurations are $A$, $B$ and $a$, $b$, respectively. Limiting our attention to incompressible materials, the deformation gradient can be written in spherical basis as
\begin{equation}
    \left[\nten{F}\right] =  \begin{bmatrix} \lam_r  & 0 & 0 \\ 0 & \lam_\theta & 0 \\ 0 & 0 & \lam_\phi \end{bmatrix} =  \begin{bmatrix} R^2/r^2  & 0 & 0 \\ 0 & r/R & 0 \\ 0 & 0 & r/R \end{bmatrix}
\end{equation}
where $\lam_{i}$  $(i = r,\theta,\phi)$ are the principal stretches associated with $\nten{F}$ and $\lam = \lam_\theta = \lam_\phi = r/R$ is the circumferential stretch. For the spherically symmetric deformation, the tensorial multiplicative decomposition \eqref{eq_F_decomp} implies the following scalar decompositions for the principal stretches
\begin{equation}
 \label{eq_lambda_relation}
\lambda_i =\lambda_{in}^e \lambda_{in}^v  \quad \text{ for}  \quad n=1,2,...,N \text{ and } i = r,\theta,\phi
\end{equation}
where $\lambda^e_{i n}$ and  $\lambda^v_{i n}$ $(i = r,\theta,\phi)$ are the principal stretches associated with $\nten{F}^e_n$ and ${\nten{F}_n^v}$, respectively. Incompressibility conditions ($\textrm{det}(\nten{F})=\textrm{det}(\nten{F}{\nten{F}_n^v}^{-1}) =1$) and symmetry imply
\begin{equation}
 \label{eq_lambda_relation_2}
    \left(\lambda_{rn}^e\right)^{-1/2} = \lambda_{\theta n}^e = \lambda_{\phi n}^e \quad,\quad \left(\lambda_{rn}^v\right)^{-1/2} = \lambda_{\theta n}^v = \lambda_{\phi n}^v (=\lambda_{n}^v)
\end{equation} Additionally, for any spherical sub-region incompressibility implies 
\begin{equation}
r^3 - a^3 = R^3 - A^3    \label{eq:incomp1}
\end{equation} 
which for the entire body reads $b^3 - a^3 = B^3 - A^3$. The circumferential stretch can then be written in terms of the circumferential stretch at the cavity wall, $\lam_a$ , as
\begin{equation}
\label{eq:lam_1}
    \lam(R,t) = \frac{r(R,t)}{R} = \left(1 + (\lam_a^3(t)-1)\left(\frac{A}{R}\right)^3\right)^{1/3}    \quad; \quad \lam_a (t) = \frac{a(t)}{A}
\end{equation}
The circumferential stretch at the outer radius of the body, $\lam_b$, is thus given as 
\begin{equation}
   \lam_b(t) = \lam(B,t) = \frac{b(t)}{B} = \left(1+\left(\lam_a^3(t) -1\right)\left(\frac{A}{B}\right)^3 \right)^{1/3} \label{eq:boundarystretches}
\end{equation}
For modelling the experiments in this paper, $B/A$ is taken to be 1000. We remark that the modelling results for an infinite solid ($B/A \to \infty$) would be indistinguishable from the results for the $B/A$ value chosen here\footnote{The material response for the finite geometry ($B/A =1000$) will deviate from that of the infinite solid at very high stretches ($\sim 100$) but such stretches would be physically impossible to realize without fracture.}, as shown in \cite{intimate_2019}. 

The radial velocity and acceleration are derived from \eqref{eq:incomp1} as
\begin{equation}
\label{eq:velacc}
\dot{r}=\frac{a^2\dot{a}}{r^2}\quad;\quad \ddot{r}= \frac{2a\dot{a}^2 + a^2\ddot{a}-2r\dot{r}^2}{r^2}
\end{equation}
Thus, the motion of the entire body is described by the motion of the inner cavity wall due to incompressibility. In what follows, we relate the motion of the cavity wall to the pressure loading $P(t)$ by using the radial equation of motion and employing the constitutive model developed in the previous subsection.

\subsection{Radial equation of motion}
\label{subsec:radialeqn}
  For the spherically symmetric deformation field, the only non-trivial equation of motion is along the radial direction
\begin{equation}
\label{eq:eqmotion}
    \pdv{\sigma_r}{r} - \frac{2s}{r} = \rho \ddot{r}
\end{equation}
 where $s=\sigma_\theta-\sigma_r$ is the difference between $\sigma_\theta =\sigma_\phi$ and $\sigma_r$, namely the circumferential and radial principal Cauchy stress components, respectively, and $\rho$ is the constant mass density. Substituting $\ddot{r}$ from \eqref{eq:velacc} in \eqref{eq:eqmotion} and integrating over the whole body from $r=a(t)$ to $r=b(t)$ yields
 \begin{subequations}\label{eq:pressure_total}
 \begin{align}
        P(t) &= S(t) + P _{in} (t) \\ 
             P _{in} (t) &=\rho A^2\left[\left(2\dot{\lam}_a^{2}+\lam_a\ddot{\lam}_a\right)\left(1-\frac{\lam_a}{\lam_b}\frac{A}{B}\right)-\frac{\dot{\lam}_a^{2}}{2}\left(1-\frac{\lam_a^{4}}{\lam_b^{4}}\frac{A^4}{B^4}\right)\right], \quad S(t)=\int_{a}^{b}\frac{2s}{r}\dd{r}
 \end{align}
\end{subequations}
where we have used the definitions for boundary stretches in \eqref{eq:lam_1}-\eqref{eq:boundarystretches} and employed the boundary conditions $\sigma_r(r=a,t)= -p_a(t)$ and $\sigma_r(r=b,t)= -p_b(t)$. Note that the gauge pressure at the cavity wall $P(t)$ differs from the gauge pressure inside the fluid in the cavity, $P_f(t)$, due to surface tension effect ($P(t) = P_f(t) - 2\gamma/a(t)$) but this has already been accounted for in the experimental pressure curves,  as seen in \cref{eq:corrected_P}. The term $S(t)$ represents the pressure due to the elastic  material resistance whereas $P _{in} (t)$ is the pressure that results from inertial effects. The inertial term, for the slow expansion rates in our experiments, is insignificant\footnote{For our experimental protocol, $\ddot{a}$ is theoretically unbounded at the end of expansion and retraction cycles but in reality the volume control by the testing machine causes smoothing of the time profile of $a$ and thus results in finite $\ddot{a}$. The inertial pressures caused by these cavity wall accelerations are still insignificant compared to pressure arising from elastic material resistance.} compared to the pressure generated due to elastic material resistance, but is nevertheless accounted for. Earlier derivation of the above relation can be found in \cite{cohen2015dynamic}. In the following subsection, we employ the constitutive model developed in \Cref{subsec:constit} to evaluate the term $S(t)$. 

\subsection{Constitutive model applied to equation of motion}

For spherically symmetric deformation, the free energy density in \eqref{free_en_total} can be rewritten in terms of principal stretches as
\begin{gather}\label{eq_free_energy}
\begin{aligned}
   \psi &= \hat{\psi}_0(\nten{F}) + \sum_{n=1}^{N} \hat{\psi}_n(\nten{F}_n^e) 
    =  \bar{\psi}_0(\lambda_r, \lambda_\theta, \lambda_\phi) + \sum_{n=1}^{N} \bar{\psi}_n (\lambda^e_{r n}, \lambda^e_{\theta n}, \lambda^e_{\phi n})
\end{aligned}
\end{gather}
which, upon substitution of  \cref{eq_lambda_relation,eq_lambda_relation_2}, can be rewritten in terms of $\lambda$ and $\lambda_n^v$,  as
\begin{gather}
\label{eq_free_energy2}
\begin{aligned}
    \psi &= \bar{\psi}_0 \left(\lambda^{-2}, \lambda, \lambda\right) + \sum_{n=1}^{N} \bar{\psi}_n \left( \left(\frac{\lambda_n^v}{\lambda}\right)^2, \frac{\lambda}{\lambda_n^v}, \frac{\lambda}{\lambda_n^v}\right)\\
        &= \tilde{\psi}_0(\lambda) + \sum_{n=1}^{N} \tilde{\psi}_n ( {\lambda}/{\lambda_n^v} )
\end{aligned}
\end{gather}
This free energy density results in the following expression for $s = \sigma_\theta-\sigma_r$ (see \ref{spherical}), 
\begin{equation}\label{eq:s_psi}
    s = \frac{\lambda}{2}{\tilde{\psi}}_0'(\lambda) + \sum_{n=1}^{N} \frac{\lambda}{2\lambda^v_n}{\tilde{\psi}}'_{n}( {\lambda}/{\lambda_n^v})
\end{equation}
Using \eqref{eq:incomp1} to employ the transformation $\dd{r}/r = \dd{\lam}/(\lam (1-\lam^3))$, we rewrite \cref{eq:pressure_total}  using \eqref{eq:s_psi} as
 \begin{equation}\label{eq:final_P}
    P(t) = \int_{\lam_a}^{\lam_b} \frac{{\tilde{\psi}}_0'(\lam)}{1-\lam^3} \dd{\lam} +  \displaystyle \sum_{n= 1}^{N} \displaystyle \int_{\lam_a}^{\lam_b} \frac{{\tilde{\psi}'_{n}(\lam}/{\lambda^v_n})}{\lambda^v_n \left(1-\lam^3\right)}   \dd{\lam} + P _{in}(t) 
\end{equation}
where $\lam_a$ and  $\lam_b$ are defined in \cref{eq:lam_1,eq:boundarystretches}. The evolution law \eqref{general_evoln_viscous} defines the viscous stretch $\lam_n^v$ implicitly through the differential equation (see \ref{spherical})
\begin{equation}
\label{Kinetic_Law_spec_gen}
 \dot{\lam}_n^v = \frac{\lam}{6 \eta_n I_{1n}^v } \tilde{\psi}_n'(\lam/{\lambda^v_n}) \qquad , \qquad I_{1n}^v = {2\lam_n^v}^2 + {\lam_n^v}^{-4}
\end{equation}
along with an initial condition for $\lam_{n}^v(R,t=0)$. 

Finally, for a prescribed deformation of the cavity wall, $\lam_a(t) = a(t)/A$, using the relations in \cref{eq:boundarystretches,eq:lam_1}, \cref{Kinetic_Law_spec_gen} can be integrated over time to evaluate the pressure variation in \cref{eq:final_P} where $P_{in}(t)$ is defined in \eqref{eq:pressure_total}. The first term in the right hand side of \cref{eq:final_P} corresponds to the pressure due to the elastic part of the response that is in thermodynamic equilibrium. The second term is the pressure that arises from the elastic part of the response that is not in thermodynamic equilibrium,
namely, the elastic part that decays in time through viscous dissipation. It remains to prescribe specific free energy functions, which are chosen next to best represent the experimental results. 

\subsection{Choice of free energy functions}

 Following previous experimental cavity expansion studies, we employ the  incompressible neo-Hookean free energy functions, which we find sufficient to capture the material response,
 \begin{equation}\label{eqb_psi}
    \hat{\psi}_0(\nten{F}) = \dfrac{E}{6}\left( I_1 -3\right) \quad, \quad  \hat{\psi}_n(\nten{F}{\nten{F}_n^v}^{-1}) = \dfrac{\alpha_n E}{6}\left( I_{1n}^e -3\right) ~ \text{ for }  n=1,2,...,N
\end{equation}
where $I_1 = \textrm{tr}\left( \nten{C} \right)$ and $I_{1n}^e = \textrm{tr}\left( \nten{C} {\nten{C}_n^v}^{-1}\right)$ with $\nten{C}=\nten{F}^T\nten{F}$ and $\nten{C}_n^v = {\nten{F}_n^v}^T \nten{F}_n^v$. The non-negative material parameter $\alpha_n$ is the ratio of the modulus of the $n^{\textrm{th}}$ non-equilibrium branch to the modulus $E$ of the equilibrium branch. In general we can choose different free energy functions for the different branches of the rheological model (\Cref{fig:rheo_model}), where the chosen free energy functions can have more material parameters than the above neo-Hookean model, which requires only one parameter (the modulus) per branch. The choice of free energy functions can have non-trivial consequences for nonlinear phenomena observed in other deformation modes, see for example \cite{chockalingam2020shear}.

 For the free energy functions in \cref{eqb_psi}, the Cauchy stress is readily derived using \eqref{cauchy_general_constit} as
\begin{equation}
\label{Cauchy_general}
    \bm{\sigma} = \frac{E}{3}\nten{F}\nten{F}^T + \sum_{n=1}^{N} \frac{\alpha_nE}{3}\nten{F}{\nten{C}_n^v}^{-1}\nten{F}^T - p\nten{I}
\end{equation}
whereas the evolution law \eqref{general_evoln_viscous}  specializes to 
\begin{equation}\label{general_evoln_viscous_cauchy}
       \nten{\dot{C}}_n^v = \frac{8}{\tau_n I_{1n}^v} \left(\nten{C} - \frac{1}{3}\left(\nten C\cdot {\nten{C}_n^v}^{-1}\right)\nten{C}_n^v \right) \quad, \quad \tau_n = \frac{12 \eta_n}{\alpha_n E} > 0 
\end{equation}
The viscous material parameter $\tau_n$ quantifies the timescale over which stress in the $n^{th}$ non-equilibrium branch decays in a stress relaxation experiment. Further, \cref{eq:final_P} simplifies to
\begin{equation}\label{eq:Pfinal2}
        P(t) = \frac{E}{6}\left(4 \lam_{b}^{-1} + \lam_{b}^{-4} - 4 \lam_{a}^{-1} -   \lam_{a}^{-4} \right) +  \displaystyle \sum_{n= 1}^{N} \displaystyle \int_{\lam_a}^{\lam_b} \frac{{\tilde{\psi}}'_{n}({\lam}/{\lambda^v_n})}{\lambda^v_n \left(1-\lam^3\right)}   \dd{\lam} + P _{in}(t) 
\end{equation}
where
\begin{equation}\label{eq:psi12}
   \tilde{\psi}_n ({\lam}/{\lambda^v_n}) = \frac{\alpha_n E}{6} \left( 2\left(\frac{\lam}{\lam_n^v}\right)^2 + \left(\frac{\lam}{\lam_n^v}\right)^{-4} -3\right) 
\end{equation}
and the evolution law \eqref{general_evoln_viscous_cauchy} defines the viscous stretch $\lam_n^v$ implicitly through the differential equation 
\begin{equation}
\label{Kinetic_Law_spec}
    \dot{\lam}_{n}^v   = \frac{4}{3 \tau_n}\frac{1}{\left({2\lam_n^v}^2 + {\lam_n^v}^{-4}\right)}\left( \frac{\lam^6 - {\lam_n^v}^6}{{\lam_n^v} \lam^4} \right)
\end{equation}
and an initial condition for $\lam_{n}^v(R,t=0)$, as described next.  

\subsection{Initial conditions}
\label{subsec:in_conds}
Integration of \eqref{Kinetic_Law_spec} for the viscous stretch $\lam_n^v(R,t)$ requires an initial condition. For our experiments, we perform separate time integrations for the expansion-relaxation and retraction-recovery part of every loading cycle. For the initial condition at the start of every expansion or retraction, we assume that the material is fully relaxed/recovered, i.e. that the contributions to the gauge pressure from the non-equilibrium branches have died out in \cref{eq:Pfinal2}. This yields the initial condition $\lam_n^v(R,t_{start}) = \lam(R,t_{start})$ where $t_{start}$ is the time when the corresponding expansion or retraction begins and $\lam(R,t_{start})$ is determined from \eqref{eq:lam_1} given $\lam_a(t_{start})$.

\section{Fitting procedure}
\label{sec:fittedresults}

Having prescribed the constitutive model, the associated material parameters remain to be estimated from the experimentally measured pressure response. By isolating the equilibrated part of the measured pressure response, the equilibrium branch modulus $E$ and the effective stress free size of the initial defect, $A$, are first estimated independent of the dynamic pressure
response. The non-equilibrium branch parameters are subsequently estimated from a best fit to the dynamic part of the measured pressure response.

\subsection{Quasistatic parameters}
\label{subsec:quasiparam}

 The equilibrated pressures at the start and end of the expansion-relaxation part of any cycle ($P_{s_1}, P_{s_2})$ are related to the equilibrium branch response, after the contributions from the non-equilibrium branches have died away, as seen from\footnote{Inertial pressure $P_{in}(t)$ also vanishes at the equilibrium states since cavity stretch $\lam_a$ is held constant.} \eqref{eq:Pfinal2} . Thus the equilibrated pressures at $(a_1,t_1)$ and $(a_2,t_2)$ are used to estimate the values of $E$ and $A$,
\begin{equation}
    P_{s_1} =  \frac{E}{6} \left(4 \lam_{b_1}^{-1} + \lam_{b_1}^{-4} - 4 \lam_{a_1}^{-1} -   \lam_{a_1}^{-4} \right) \quad,\quad    P_{s_2} =  \frac{E}{6} \left(4 \lam_{b_2}^{-1} + \lam_{b_2}^{-4} - 4 \lam_{a_2}^{-1} -   \lam_{a_2}^{-4} \right) 
\end{equation}
where $\lam_{a_i} = a_i/A \ (i=1,2)$ and $\lam_{b_i}$ is related to $\lam_{a_i}$ using \eqref{eq:boundarystretches}. The estimated values of $E$ and $A$ from the equilibrated pressures of each loading cycle, for the representative sample S50-00,  can be seen to be approximately constant in  \Cref{Table1_allcycles_params} (second and third columns). This is a consequence of the fact that the experimental equilibrated pressures do not vary much across loading cycles. The slight but steady increase in the value of $A$ across cycles is attributed to likely fatigue/damage, as discussed in \Cref{sec:Modelresult}. We note here that while our estimated equilibrium elastic modulus is about $15~\textrm{kPa}$, the regular VCCE technique reports a significantly higher modulus of about $75~\textrm{kPa}$ and the Cavitation Rheology technique reports a modulus of about $25~\textrm{kPa}$, for the same material composition \citep{vcce-raayai}. One source of discrepancy is the fact that we have eliminated the Mullins effect which results in some softening. However, a greater discrepancy arises from the choice of loading protocol as elaborated in \Cref{sec:constQ}.

\subsection{Dynamic parameters}

The remaining non-equilibrium material parameters are estimated using a nonlinear least squares fitting method that employs a material subroutine to integrate \cref{eq:Pfinal2,Kinetic_Law_spec}. The integration and fitting procedure are described in more detail in \ref{app:constit}. While our experimental cavity expansion protocol (\Cref{fig:model_result}) includes multiple cycles of expansion and relaxation followed by retraction and recovery, parameter fitting can either be optimized for the entire time of the experiment or for specific sub-intervals. In this work, we conduct the fitting on the expansion-relaxation part of the material response\footnote{There is no possibility of separation of syringe plunger and fluid during expansion-relaxation and as such the pressure data can be regarded to be more reliable than for retraction-recovery.}. Nonetheless, as a validation, the fitted parameters will be seen to also capture the retraction-recovery profiles well. For our experimental results, we find that  two non-equilibrium branches ($N=2$) and thus four non-equilibrium parameters ($\alpha_1, \alpha_2, \tau_1, \tau_2$) are sufficient to capture the experimentally measured material response. While a different choice of the kinetic evolution law, instead of \cref{general_evoln_viscous_cauchy},  could possibly capture the experimental response well using a single non-equilibrium branch, there is no straightforward way to determine such an optimal kinetic law. Among the different kinetic laws we surveyed, the chosen evolution law in \cref{general_evoln_viscous_cauchy} performs the best in capturing the experimental results.

Representative fitted parameters obtained by individually fitting to the expansion-relaxation part of each loading cycle of the experimental curves for sample S50-00, are shown in \Cref{Table1_allcycles_params}. It is seen that the estimated non-equilibrium branch parameters exhibit a moderate dependence on the expansion rate. This dependence could potentially be reduced by optimizing the kinetic evolution law, by either varying the free energy functions of the non-equilibrium  branches ($\psi_n$), or by choosing a different form of the evolution law \eqref{Kinetic_Law_spec_gen}. Nonetheless, any fitted model can only be optimized to perform within the range of expansion rates probed in the experiment. In this work, to obtain a best fit for the entire range of expansion rates across all loading cycles using a single set of parameters, we examine the error associated with the fitted dynamic parameters of the individual cycles, when used to predict the pressure for all loading cycles while using the values of $E$ and $A$ fitted for the corresponding cycles\footnote{This is because we assume that there is slight fatigue between cycles that is not accounted for in the material model and the variation in the fitted values for $E$ and $A$ across the loading cycles is small.}. 
 \begin{table}
  \centering \footnotesize
 \caption{Fitting parameters for representative sample S50-00 (see \Cref{sec:Modelresult,sec:Oil_expts}).} \label{Table1_allcycles_params}
\begin{tabular}{cccccccc}
\hline 
Fitting Cycle &$\dot{a}$ (\si{\mm\per\s})& $A$ (\si{mm}) & $E$ (\si{kPa})  & $\alpha_1$ &   $\tau_1$ (\si{s}) & $\alpha_2$ & $\tau_2 $ (\si{s})  \\ 
\hline 
1 & 0.02 & 0.353 & 14.71 & 0.597 & 2.76 & 0.097 & 154.63 \\ 
2 & 0.04 & 0.356 & 14.77 & 0.789 & 1.54 & 0.109 & 129.07 \\ 
3 & 0.08 & 0.358 & 14.77 & 0.927 & 1.00 & 0.116 & 123.18 \\ 
4 & 0.16 & 0.360 & 14.79 & 1.246 & 0.56 & 0.128 & 110.56 \\ 
5 & 0.32 & 0.363 & 14.83 & 1.599 & 0.32 & 0.140 & 97.16 \\ 
\hline 
\end{tabular}
 \end{table}

 To evaluate the fit of a given set of parameters on a single loading cycle, we define the following errors
\begin{equation}
\label{eq:errors}
    \epsilon = \sqrt{ \frac{\int_{t_1}^{t_3}(P-P_{e})^2 \dd{t_{rel}}}{\int_{t_1}^{t_3}P_{e}^2 \dd{t_{rel}}} }, ~~     \epsilon_{exp} = \sqrt{ \frac{\int_{t_1}^{t_2}(P-P_{e})^2 \dd{t_{rel}}}{\int_{t_1}^{t_2}P_{e}^2 \dd{t_{rel}}} },~~  \epsilon_{max} = \underset{t_{rel} \in [t_1 t_3]}{\textrm{max}} ~\left(\abs{\frac{P - P_{e}}{P_{e}}}\right)
\end{equation}
where $P$ is the gauge pressure predicted by the viscoelastic material model for the given set of parameters and $P_{e}$ is the experimentally measured gauge pressure\footnote{Note that for only this section, we differentiate the experimental and theoretical gauge pressures with different symbols.}. The fitting error for the time of the entire expansion-relaxation is estimated by $\epsilon$ whereas $\epsilon_{exp}$ denotes the fitting error on the expansion part of the loading cycle alone. The integrals in \eqref{eq:errors} are evaluated numerically using a trapezoidal rule since the pressure data is discrete. 
To evaluate the fit over all loading cycles for a given set of material parameters we define the following overall errors
\begin{equation}
    \epsilon^{tot} = \underset{\textrm{all cycles}}{\textrm{mean}}\ \epsilon \quad,\quad \epsilon_{exp}^{tot} = \underset{\textrm{all cycles}}{\textrm{mean}}\ \epsilon_{exp}\quad,\quad \epsilon_{max}^{tot} = \max_{\textrm{all cycles}} \epsilon_{max} 
\end{equation}
 \begin{table}[!htb]
  \centering \footnotesize
 \caption{Errors in the pressure predicted by the viscoelastic material model, for all loading cycles, using dynamic material parameters fitted on Cycle 3 (third row in \Cref{Table1_allcycles_params}), for representative sample S50-00.} \label{Table2_fit_Eval}
\begin{tabular}{ccccc}
\hline 
Cycle &$\dot{a}$ (\si{mm/s})& $\epsilon$ (\%) & $\epsilon_{exp}$ (\%)  & $\epsilon_{max}$ (\%) \\ 
\hline 
1 & 0.02 & 0.32 & 1.22 & 1.76 \\ 
2 & 0.04 & 0.33 & 1.01 & 2.12 \\ 
3 & 0.08 & 0.36 & 0.47 & 2.67 \\ 
4 & 0.16 & 0.38 & 1.43 & 4.02 \\ 
5 & 0.32 & 0.43 & 2.78 & 6.75 \\ 
\hline 
&  & $\epsilon^{tot}(\%) = $ 0.36 & $\epsilon_{exp}^{tot}$ (\%) = 1.38 & $\epsilon_{max}^{tot}$ (\%) = 6.75 \\ 
\hline 
\end{tabular}
 \end{table}
\begin{figure}[!h]
    \centering
    \includegraphics[width = \textwidth]{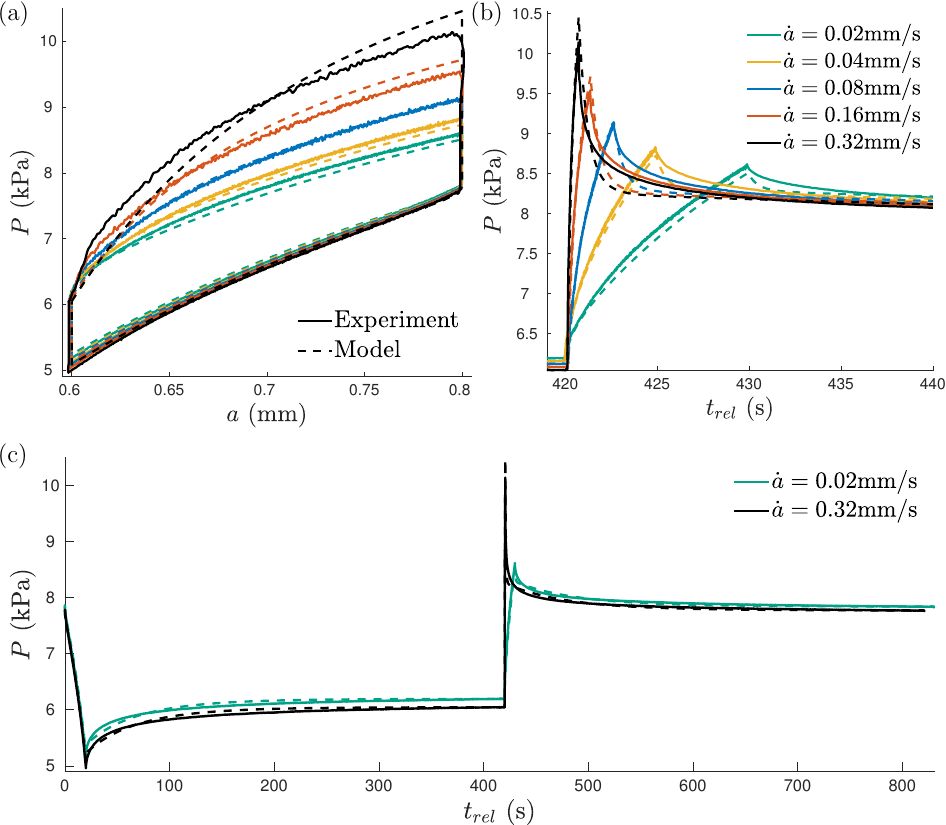}
    \caption{Comparison of the pressure response predicted by the viscoelastic material model with the experimental pressure data. The non-equilibrium material parameters fitted to loading Cycle 3 (\Cref{Table1_allcycles_params}) have been used to predict the pressure response for all loading cycles. The responses for only the extreme loading rates are shown in (c) to avoid clutter.}
    \label{fig:fitted_model_Result}
\end{figure}
The material parameters that give the lowest overall errors are reported as the optimal fitted parameters for the material\footnote{For reporting optimal values for $E$ and $A$, we use their fitted values from the cycle whose fitted dynamic parameters give the lowest overall errors.}. For the representative sample S50-00, the dynamic parameters fitted on Cycle 3 yield the lowest overall errors for pressure prediction across all loading cycles. The prediction errors for individual cycles using these optimal parameters are reported in \Cref{Table2_fit_Eval}. It can be seen that a single set of parameters can accurately predict the pressure response over the entire expansion-relaxation process for all loading cycles (stretch rates spanning $\num{e-2}$ - $1$ $\textrm{s}^{-1}$), with the maximum absolute relative error always less than $7\%$. The predicted material response using these parameters is visually compared with the experimental curves in \Cref{fig:fitted_model_Result}, where it is seen that the retraction-recovery part of the response is also well predicted.

\subsection{A cautionary note on  viscoelastic effects in constant volumetric rate expansion}
\label{sec:constQ}
\begin{figure}[!h]
    \centering
    \includegraphics[width=\textwidth]{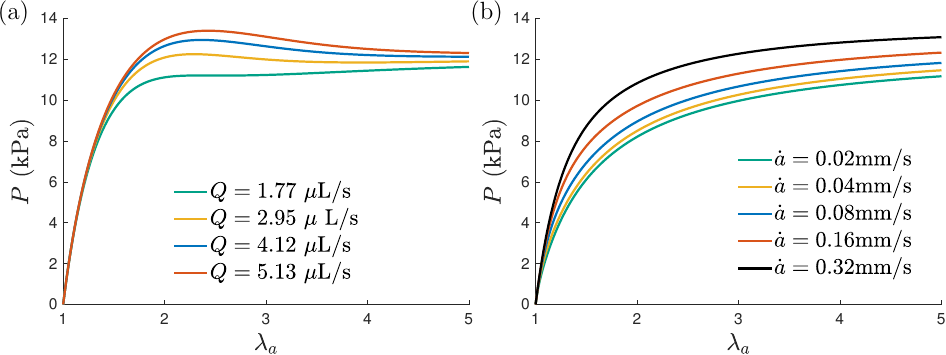}
    \caption{Pressure versus cavity stretch response predicted by the optimal fitted viscoelastic material parameters for sample S50-00, for different loading scenarios. (a) The volume expansion rate $Q$ is constant. There is not much rate dependence observed during the expansion until high stretches are accessed and the pressure can drop for high stretch values. (b) The cavity stretch rate $\dot{\lam}_a = \dot{a}/A$ is constant. Rate dependence can be clearly seen and there is no pressure drop even at high stretches.}
    \label{fig:constQ_prediction}
\end{figure}

We now use the fitted material parameters from the previous section to explain both the lack of rate dependence observed in the pressure response in the VCCE experiments of \cite{vcce-raayai}, as well as the significantly higher modulus values reported therein. Recall that the main difference between the experimental protocol in \cite{vcce-raayai} compared with the present work, is the expansion process. Here we consider constant rate radial expansion (i.e. $\dot{a} =$ const) in contrast to constant volume expansion rate (i.e. $Q=4\pi a^2 \dot a =$const), as discussed in \Cref{sec_final_protocol}. 

Using the optimal fitted material parameters (third row of \Cref{Table1_allcycles_params}) for sample S50-00, we perform numerical integration to predict the pressure response for the case of constant volume expansion rate $Q$, starting at an undeformed cavity size $A$ at $t=0$. Accordingly, the cavity size as a function of time is given by
\begin{equation}
    a(t) = \left(\frac{3 Q t}{4 \pi}+A^3\right)^{1/3}
\end{equation}
We consider $A=0.35~\textrm{mm}$ and use the four volumetric rates reported in \cite{vcce-raayai}, namely $Q = 1.77,~ 2.95 ,~4.12 ,~5.13 ~\rm{\mu L /s}$, to numerically evaluate the pressure response. The resulting pressure versus cavity stretch curves are plotted in \Cref{fig:constQ_prediction}(a). Almost no rate dependence is seen among the pressure curves for the four different seemingly small volumetric rates, at least until larger stretches are accessed. This lack of rate dependence is consistent with the experimental observations of \cite{vcce-raayai}. For comparison, the integrated results for the case of constant cavity stretch rate loadings  ($a = \dot{a} t$) are shown in \Cref{fig:constQ_prediction}(b), for the same material parameters with the expansion starting from the same undeformed cavity size $A=0.35~\rm{mm}$. The rate dependent material response can be easily seen from these curves throughout the expansion. As explained earlier in \Cref{sec_final_protocol}, if the initial cavity size is sufficiently small in a spherical expansion setting, even small volumetric rates can result in high stretch rates initially. This explains the apparent saturation of the dynamic material response observed in \Cref{fig:constQ_prediction}(a).

To further understand this response and its effect on the measurement of the elastic properties, we can write the instantaneous modulus, $E_{inst}$, for the rheological model in \Cref{fig:rheo_model} with $N=2$ and with the free energy functions chosen in \cref{eqb_psi},  as
\begin{equation}
\label{eq:dynamic_modulus}
    E_{inst} = E (1+\alpha_1 + \alpha_2)
\end{equation}
This is the effective dynamic modulus of the material before onset of appreciable viscoelastic effects.  Given the rate-insensitive response shown for moderate stretch levels in \Cref{fig:constQ_prediction}(a), it is expected that a purely hyperelastic fit (only equilibrium branch free energy) to the expansion profile would yield the instantaneous modulus, which according to the fitted values for $\alpha_1$ and $\alpha_2$ in \Cref{Table1_allcycles_params} can over estimate the equilibrium modulus $E$ by more than 250 $\%$.  This would in part explain the higher modulus values reported for the same PDMS composition in \cite{vcce-raayai} and emphasizes the need for stretch rate controlled VCCE experiments that can account for viscoelastic effects to accurately capture the material parameters.  

An additional feature of the constant volumetric rate expansion is the overshot of the pressure, as observed by the appearance of peak values  in \Cref{fig:constQ_prediction}(a).
Although the cavity continues to expand, the stretch rate decreases with increasing expansion ($\dot{a} = Q/(4 \pi a^2)$), thus the material response transitions from the stiffer, instantaneous, behavior characterised by the instantaneous modulus $E_{inst}$, to the softer behavior characterised by the equilibrium modulus $E$. Note that this transition is radially dependent, and the viscoelastic stiffening is most noticeable in the near vicinity of the cavity. 


For the experiments in \cite{intimate_2019}, fracture, characterised by sharp sudden drop in pressure as opposed to the smooth and slight decrease in \Cref{fig:constQ_prediction}(a), is observed at such high stretches. For the constant cavity stretch rate expansion the pressure does not drop even for high stretches, implying that a pressure drop would have to arise solely from fracture\footnote{The pressure can drop even for constant stretch rate loadings at very large stretches, however the material would fracture much earlier in reality.}, thus making it an even more attractive choice of loading protocol. 

\section{Characterising the viscoelastic response of tuneable PDMS samples}
\label{sec:Oil_expts}
Having established the experimental method and the fitting of the experimental results to a generalized viscoelastic material model, we now apply our technique to study the viscoelastic response of soft PDMS rubber samples with tuneable content of non-reactive Silicone (PDMS) oil. This approach is inspired from the fracture study of such a system by \cite{shelby_oil}. Here we aim to examine the ability of our technique to capture changes in the constitutive response, which in this case are due to the oil content. 

\subsection{Sample preparation}
The Sylgard 184 (Dow Corning) PDMS base is diluted with different weight percent of non-reactive PDMS oil ($\mathrm{\mu}$MicroLuburol - 350cSt). The diluted PDMS mixture is then mixed with cross-linker to obtain different PDMS to cross-linker mass ratios (PDMS:CL), as defined in \Cref{tab_samples}.  Note that,
in contrast to the fabrication procedure in \cite{shelby_oil}, where
a base and cross-linker mix with a given mass ratio is subsequently diluted with oil, here we conserve the total mass fraction of the cross-linker to PDMS content including both the  base and non-reactive oil.
The final mixture of base, oil, and cross-linker, is homogenized in two cycles in a planetary centrifugal mixer. The mix is poured into plastic sample cups (dimensions - $\varnothing$ 2.5'' x 1.75'' height) and subsequently degassed in a desiccator for approximately 1 hour. The samples are cured in an oven at 100\si{\celsius} for 2 hours and then left to cool at room temperature. All tests are performed 2 - 3 weeks after cure.

\begin{table}[!htb]
\caption[Composition of Diluted PDMS Samples]{Sample compositions. The Sylgard 184 PDMS base is diluted with non-reactive Silicone (PDMS) oil. The diluted PDMS mixture is mixed with cross-linker to obtain different PDMS mix to cross-linker mass ratios (PDMS:CL). }\label{tab_samples}
\centering\footnotesize
\begin{tabular}{c|cc|ccc}
\hline
Sample Reference&\multicolumn{2}{c|}{Composition}&\multicolumn{3}{c}{Effective Mass Ratio}\\ 
&PDMS:CL &Oil &  Base & Oil & Cross-linker   \\
\hline\hline
S50-00&50:1& 0\%& 50&0&1\\
S50-10&50:1& 10\% & 45&5&1\\
S50-20&50:1& 20\% & 40 & 10&1\\
\hline         
S48-00&48:1 & 0\% & 48&0&1\\
S48-10&48:1&10\% &  43.2&4.8&1\\
S48-20&47.3:1&20\% & 37.8&9.5&1\\
\hline
\end{tabular}
\end{table}

\subsection{Results}
\begin{figure}[!htb]
    \centering
    \includegraphics[width=\textwidth]{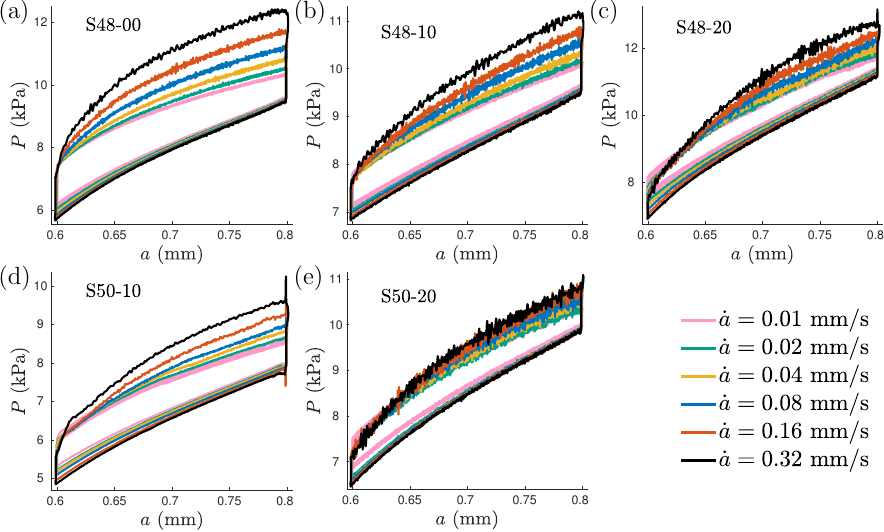}
    \caption{Experimental pressure profiles for the tuneable PDMS-Silicone oil system discussed in \Cref{sec:Oil_expts}. Compositions for the different samples are listed in \Cref{tab_samples}. The profile for sample S50-00 is shown in \Cref{fig:model_result}. Increasing oil content can be seen to reduce the rate dependence of the material response.  }
    \label{fig:all_samples}
\end{figure}

\begin{figure}
    \centering
    \includegraphics[width=\textwidth]{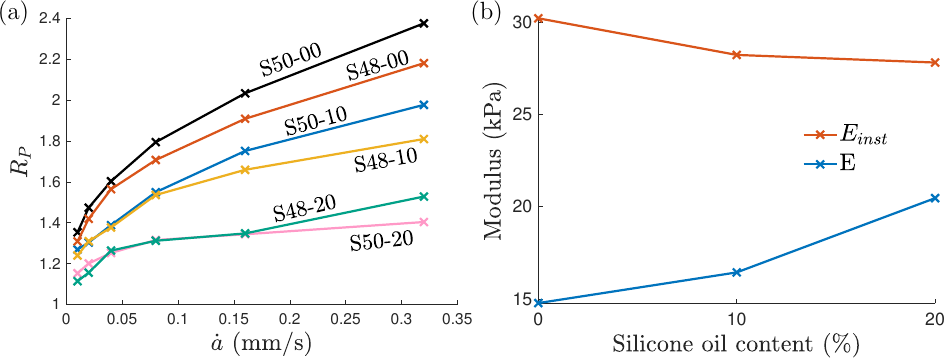} 
    \caption{(a) Dynamic amplification ratio $R_P$ increases with cavity expansion rates for all samples and decreases with increasing Silicone oil content for a given cross-linker mass fraction. (b) Plots of the modulus of equilibrium branch, $E$, and the instantaneous modulus, $E_{inst} = E ( 1+ \alpha_1 +\alpha_2)$, for fixed cross-linker mass fraction of $1/50$. The two moduli approach each other with increasing Silicone oil content.}
    \label{fig:proof_trends}
\end{figure}

The full experimental protocol described in \Cref{tab_fullp} was carried out for all the material samples listed in \Cref{tab_samples}. The resulting pressure profiles are shown in \Cref{fig:all_samples}. It can clearly be seen that increasing Silicone oil content, for a given cross-linker mass fraction, leads to decreasing rate dependence of the material response. Plots of the dynamic amplification ratio, defined in \cref{eq:DAR}, are shown in \Cref{fig:proof_trends}(a). The ratio increases with cavity expansion rates for all samples, indicating that the dynamic material stiffening (compared to the quasistatic response) is higher at higher stretch rates. The ratio also decreases with increasing Silicone oil content, for a given cross-linker mass fraction, at all expansion rates. This is consistent with the observation of decreasing rate dependence with increasing Silicone oil content, for a given cross-linker mass fraction.  We also note that the equilibrated pressures show more variation across the loading cycles with increasing Silicone oil content, with the largest variation of about $1.5~\textrm{kPa}$. This could be an indicator of possibly more pronounced fatigue effects with increasing oil content.

 \begin{table}
  \centering \footnotesize
 \caption{Optimal fitted material parameters for all samples listed in \Cref{tab_samples}. The parameter values from the fitting cycle that gives the best fit for all cycles is shown here, along with the total fitting errors. Parameters fitted to individual cycles for every sample can be found in \ref{app:params}. } \label{Table3_allsample_params}
\begin{tabular}{cccccccccc}
\hline 
Sample  & $A$ (\si{mm}) & $E$ (\si{kPa}) & $\alpha_1$ &$\tau_1$ (\si{s})  & $\alpha_2$  & $\tau_2 $ (\si{s})  & $\epsilon_{exp}^{tot}(\%)$ & $\epsilon^{tot}(\%)$ & $\epsilon_{max}^{tot}(\%)$ \\ 
\hline 
S50-00 & 0.358 & 14.77 & 0.927 & 1.00 & 0.116 & 123.18 & 1.38 & 0.36 & 6.75 \\ 
S50-10 & 0.406 & 16.44 & 0.637 & 1.03 & 0.079 & 156.72 & 1.41 & 0.34 & 8.40 \\ 
S50-20 & 0.403 & 20.46 & 0.315 & 1.65 & 0.044 & 139.70 & 1.69 & 0.18 & 7.52 \\ 
S48-00 & 0.382 & 18.91 & 0.899 & 1.39 & 0.102 & 135.28 & 2.25 & 0.38 & 10.58 \\ 
S48-10 & 0.355 & 17.98 & 0.380 & 1.27 & 0.070 & 131.88 & 0.93 & 0.22 & 4.48 \\ 
S48-20 & 0.428 & 24.33 & 0.505 & 0.94 & 0.041 & 172.33 & 1.36 & 0.20 & 6.27 \\ 
\hline 
\end{tabular}
 \end{table}

The optimal fitted material parameters for all the samples along with the prediction errors are recorded in \Cref{Table3_allsample_params}. For the chosen evolution law \eqref{Kinetic_Law_spec}, two non-equilibrium branches were necessary to capture the experimentally observed material response, and the fitted parameters for $\tau_1$ and $\tau_2$ indicate the presence of two distinct timescales of relaxation. The parameters fitted to individual cycles, for all samples, are listed in \ref{app:params}. The effective size of the stress free cavity, $A$, does not vary much across all samples tested here, implying that all the samples were tested in similar stretch ranges ($\sim 1.3 - 2.3$). Plots of $E$ and $E_{inst}$ (defined in \cref{eq:dynamic_modulus}) versus Silicone oil content, for the fixed cross-linker mass fraction of $1/50$, are shown in \Cref{fig:proof_trends}(b). It can be seen that the two moduli approach each other with increasing oil content. The closer the values of $E$ and $E_{inst}$, the lesser the allowance for rate stiffening in the expansion response\footnote{Note that the decreasing value of the ratio $E_{inst}/E$ with increasing Silicone oil content does not automatically imply that the rate dependence of material response is decreasing with increasing oil content, since the viscosity material parameters are not constant across samples. However, the low fitting errors in \Cref{tab_fullp} confirm that the model is capturing the experimentally observed decrease in rate dependence with increasing oil content.}. A small increase in the equilibrium branch modulus $E$ is seen with increasing oil content. 
Note that comparison of individual non-equilibrium branch parameters across samples needs to done with caution as the fitting optimisation is not unique for a multiple branch nonlinear viscoelastic model. Furthermore, the fitted parameters across different loading cycles show variation for even a given sample (\ref{app:params}). Nevertheless, for all samples, the single set of optimal parameters reported in \Cref{Table3_allsample_params} perform well across all loading rates as indicated by the low prediction errors. The fitted material parameters can be used to predict the material response for any given deformation by use of \cref{Cauchy_general,general_evoln_viscous_cauchy} along with initial conditions for the viscous deformation tensors. The constitutive model could also be coded as a user material subroutine in finite element programs that can be used with the fitted material parameters to solve boundary value problems.

\section{Summary and conclusions} 
\label{sec:Conclusions}

The VCCE technique extracts local nonlinear elastic properties in soft material samples by performing volume controlled expansion of a cavity through injection of an incompressible and immiscible fluid, combined with active monitoring of pressure inside the cavity. In this work,
several enhancements have been introduced to the VCCE technique that extend its capability to measurement of local nonlinear viscoelastic properties of soft solids at low to medium stretch rates ($\num{e-2}$ - $1$ s${}^{-1}$). First, the accuracy of the pressure measurement is vastly improved by measuring the pressure through a pressure sensor, instead of through reaction forces measured by the mechanical testing machine. For fluid injection, the change from regular syringes to gas-tight syringes of smaller cross sectional area, results in significantly more precise volume control of the cavity. The most significant modification is the new cavity expansion protocol proposed here. Instead of the constant volumetric expansion rate, conventionally used in VCCE and other needle based cavity expansion techniques, the new protocol specifies constant cavity stretch rate expansion that allows for observation of appreciable rate sensitivity. Accordingly, after eliminating the Mullins effect, several cycles of expansion-relaxation and retraction-recovery are performed at different expansion rates to capture the corresponding pressure response. 
Material parameters are determined by comparing the experimentally observed pressure profiles with theoretical predictions that are obtained using a generalized large deformation nonlinear viscoelastic model. It is shown that the equilibrium modulus can be directly inferred from the repeatable equilibrated pressures at two different cavity sizes, while the non-equilibrium parameters are determined by a best fit to the expansion-relaxation data.  

Application of the  technique  to characterize the viscoelastic response of soft PDMS samples with tuneable content of Silicone oil, shows sensitivity and repeatability.  The rate dependence of the material response is seen to reduce with increasing oil content for a given mass fraction of cross-linker. For the kinetic evolution law considered here, the fitted material parameters for all material compositions indicate two distinct timescales of relaxation, of the orders of $\sim 1$ s and $\sim 100$ s. Also, a single set of material parameters is shown to capture the pressure response across all the different cavity expansion rates with high accuracy. 

Provided the measured viscoelastic properties of soft PDMS samples, obtained in this work, we set out to explain earlier reports of rate insensitivity exhibited in needle based cavity expansion methods. This investigation elucidates the significance of the specific expansion protocol and leads to a cautionary note on the quasistatic assumption in earlier work.  It is found that even slow volumetric rate fluid injection can lead to high cavity stretch rates. These high stretch rates, which are more pronounced for smaller initial cavity size, can lead to viscoelastic stiffening that  dominates the pressure response throughout the expansion, thus leading to an illusion of rate insensitivity. This effect leads to an overestimation of the elastic modulus in earlier studies and hence, the loading protocol proposed in this paper becomes essential for accurate estimation of both quasistatic and dynamic material parameters.

This work is not without limitations. Additional advancements are needed to reliably apply this technique to more complex heterogeneous materials. Challenges might appear in isolation of a fracture free range and in interpretation of the pressure data. Additionally, the current approach assumes the material to be incompressible, and the specimen to be sufficiently large in comparison with the needle diameter. If these assumptions are relaxed, additional parameters should enter the theoretical prediction of the pressure response. 
Finally, this technique could potentially be extended, in the future, to characterise additional properties of soft materials. Fatigue that is observed across loading cycles in our experiments could perhaps be better characterised using a viscoplastic material model. Mullins effect is also captured here, for the first time in a spherically symmetric deformation setting. The VCCE technique can thus potentially aid in better characterization of this effect. 
Since the proposed technique can extract local nonlinear viscoelastic material properties in soft opaque materials while being minimally invasive, it should be a promising candidate for \textit{in vivo} viscoelastic testing of biological tissues.

\section*{Acknowledgements}         
 The authors wish to acknowledge the support of: the Army Research
Office, United States of America and Dr. Ralph A. Anthenien,
Program Manager, under award number W911NF-19-1-0275; the Office of Naval Research, United States of America and Dr. Timothy B. Bentley, Program Manager, under award number N00014-20-1-2561; and helpful conversations with Aditya Kumar (University of Illinois at Urbana-Champaign), as well as the help from Seethalakshmi (Texas A\&M University) in generation of figures.       
\appendix 

\section{Viscoelastic modelling}
\label{app:constit}

\subsection{Constitutive model}
\label{app:constit100}

Following the constitutive modelling approach in \cite{kumar2016two,kumar2017some,ghosh2020two}, kinetic evolution laws for the viscous deformation $\nten{F}_n^v$ are prescribed using dissipation potentials $\phi_n(\nten{F},\nten{F}^v_n,\nten{F}^v_n)$ as 
\begin{equation}\label{evoln_general}
    \pdv{\hat{\psi}_n}{\nten{F}^v_n}\left(\nten{F}{\nten{F}^v_n}^{-1}\right) + \pdv{\phi_n}{\nten{\dot{F}}^v_n}\left(\nten{F},\nten{F}^v_n,\nten{\dot{F}}^v_n\right) = \nten{0} \qquad \text{ for }  n=1,2,...,N
\end{equation}
The second law of thermodynamics, for isothermal processes, imposes the following constraint on the dissipation potentials
\begin{equation}
\label{dissip_ineq}
   \sum_{n=1}^{N} \left[\pdv{\phi_n}{\nten{\dot{F}}^v_n}\left(\nten{F},\nten{F}^v_n,\nten{\dot{F}}^v_n\right)\right]\cdot \nten{\dot{F}}^v_n \ge 0 
\end{equation}
for arbitrary deformation gradients $\nten{F}$, $\nten{F}^v_n$, with equality holding only when $\nten{\dot{F}}^v_n = \nten{0}$ (for $n=1,2,...,N$). The inequality \eqref{dissip_ineq} is automatically enforced if following $N$ inequalities are satisfied
\begin{equation}
   \label{dissip_ineq2}
   \left[\pdv{\phi_n}{\nten{\dot{F}}^v_n}\left(\nten{F},\nten{F}^v_n,\nten{\dot{F}}^v_n\right)\right]\cdot \nten{\dot{F}}^v_n \ge 0  \quad \textrm{for} \quad n = 1,2,...,N
\end{equation}
Extending the modelling approach in \cite{ghosh2020two} to multiple non-equilibrium branches, we choose the following dissipation potentials that satisfy \eqref{dissip_ineq2},
\begin{equation}\label{phi_general}
\phi_n\left(\nten{F},\nten{F}^v_n,\nten{\dot{F}}^v_n\right) =  \frac{1}{2} \nten{\dot{F}}^v_n{\nten{F}_n^v}^{-1}\cdot \left[\bm{\mathcal{A}}_n \left(\nten{\dot{F}}^v_n{\nten{F}_n^v}^{-1}\right)\right] \quad, \quad \bm{\mathcal{A}}_n = 2 \eta_{K_n}(I_{1n}^v) \bm{\mathcal{K}} + 3 \eta_{J} \bm{\mathcal{J}}
\end{equation}
where $I_{1n}^v = \textrm{tr}(\nten{C}^v_n)$, $\mathcal{K}_{ijkl}=1/2\left[\delta_{ik}\delta_{jl} + \delta_{il}\delta_{jk} - 2/3\    \delta_{ij}\delta_{kl} \right]$, $\mathcal{J}_{ijkl}= 1/3\ \delta_{ij}\delta_{kl}$ and $\eta_{J} \to + \infty$. We note that $\bm{\mathcal{K}}$ is the symmetric deviatoric projection tensor such that\footnote{$\textrm{symdev}(\nten{X}) = \frac{1}{2}\left(\textrm{dev}(\nten{X}) + (\textrm{dev}(\nten{X}))^T\right)$ where $\textrm{dev}(\nten{X}) = \nten{X} - \frac{1}{3}\textrm{tr}(\nten{X})\nten{I}$.} $\bm{\mathcal{K}} \nten{X} = \textrm{symdev}(\nten{X})$ and $\bm{\mathcal{J}}$ is the spherical projection tensor such that   $\bm{\mathcal{J}} \nten{X} = \frac{1}{3}\textrm{tr}(\nten{X})\nten{I}$, for any second order tensor $\nten{X}$. We select the viscosity function $\eta_{K_n}(I_{1n}^v)$ to be the following increasing function of deformation
\begin{equation}
\label{eq_app00}
    \eta_{K_n}(I_{1n}^v) = \frac{1}{2} \eta_n I_{1n}^v 
\end{equation}
where $\eta_n>0$. Differentiating \eqref{phi_general} with respect to $\nten{\dot{F}}^v_n$, we have
\begin{equation}
\label{eq_app1}
    \pdv{\phi_n}{\nten{\dot{F}}^v_n} =  \left(\bm{\mathcal{A}}_n\left(\nten{\dot{F}}^v_n{\nten{F}_n^v}^{-1}\right)\right){\nten{F}_n^v}^{-T}  
\end{equation}
Using \eqref{eq_F_decomp}, we can write
\begin{equation}
\label{eq_app2}
 \pdv{{\psi}_n}{\nten{F}^v_n} = - \nten{M}_n {\nten{F}_n^v}^{-T} \quad \textrm{where} \quad \nten{M}_n = {\nten{F}_n^e}^{T}\pdv{\hat{\psi}_n}{\nten{F}^e_n}
\end{equation}
Then, by using \cref{evoln_general,eq_app1,eq_app2}, we can write
\begin{equation}
  \label{eq_app3} \nten{M}_n =  \bm{\mathcal{A}}_n\left( \nten{\dot{F}}^v_n{\nten{F}_n^v}^{-1}\right)
\end{equation}
where we recall $\bm{\mathcal{A}}_n = 2 \eta_{K_n}(I_{1n}^v) \bm{\mathcal{K}} + 3 \eta_{J} \bm{\mathcal{J}}$. We have $\eta_J \to \infty$ which forces $\textrm{tr}(\nten{\dot{F}}^v_n{\nten{F}_n^v}^{-1})=0$, and then taking the deviatoric part\footnote{The product $\eta_J~ \textrm{tr}(\nten{\dot{F}}^v_n{\nten{F}_n^v}^{-1})$ can have a finite limit and thus we remove the spherical component from both sides of \eqref{eq_app3}.} of the two sides of \eqref{eq_app3} gives us
\begin{equation}
\label{eq_app4}
    \nten{M}_n - \frac{1}{3}\textrm{tr}(\nten{M}_n)\nten{I} = 2 \eta_{K_n}\ \textrm{sym}\left(\nten{\dot{F}}^v_n{\nten{F}_n^v}^{-1}\right)
\end{equation}
Pre-multiplying both sides of \eqref{eq_app4} by ${\nten{{F}}^v_n}^{T}$ and post-multiplying by ${\nten{{F}}^v_n}$, we get
\begin{equation}
\label{eq_app6}
   {\nten{{F}}^v_n}^{T} \nten{M}_n \nten{{F}}^v_n - \frac{1}{3}\textrm{tr}(\nten{M}_n)\nten{C}_n^v= 2 \eta_{K_n} ~\textrm{sym}\left({\nten{{F}}^v_n}^{T}\nten{\dot{F}}^v_n\right) = \eta_{K_n}  \nten{\dot{C}}_n^v
\end{equation}
which along with \eqref{eq_app00} yields the evolution equation \eqref{general_evoln_viscous}. For the free energy functions in \eqref{eqb_psi}, we get
\begin{equation}
\label{eq_app5}
    \nten{M}_n = \frac{\alpha_n E}{3} \nten{C}_n^e\quad,\quad \textrm{tr}(\nten{M}_n) = \frac{\alpha_n E}{3} \nten C\cdot {\nten{C}_n^v}^{-1}
\end{equation}
where $\nten{C}_n^e = {\nten{F}_n^e}^{T}{\nten{F}_n^e}$, and we have used $\textrm{tr}(\nten{C}_n^e) = \textrm{tr}(\nten{C} {\nten{C}_n^v}^{-1}) = \nten C\cdot {\nten{C}_n^v}^{-1}$. 
Substituting \eqref{eq_app5} in \eqref{eq_app6} and using \eqref{eq_app00}, we obtain the evolution equation \eqref{general_evoln_viscous_cauchy}.

\subsection{Spherically symmetric deformation}
\label{spherical}
For a free energy function of the form in \Cref{eq_free_energy}, the expression for the Cauchy stress in \eqref{cauchy_general_constit} yields
\begin{equation}
\label{app:cauchy_constit}
    \bm{\sigma} = \pdv{\hat{\psi}_0}{\nten{F}} \nten{F}^T + \sum_{n=1}^{N} \pdv{\hat{\psi}_n}{\nten{F}_n^e} ~ {\nten{F}_n^e}^T- p \nten{I}
\end{equation}
With the free energy density written in terms of the principal stretch components ($\bar{\psi}_0(\lambda_r, \lambda_\theta, \lambda_\phi)$ and $\bar{\psi}_n (\lambda^e_{r n}, \lambda^e_{\theta n}, \lambda^e_{\phi n})$) for the spherically symmetric deformation, the principal stress difference resulting from \eqref{app:cauchy_constit} can be shown to be 
\begin{subequations}
\begin{align}
    \sigma_\theta - \sigma_r &= \lam_\theta \pdv{\bar{\psi}_0}{\lam_\theta} -  \lam_r \pdv{\bar{\psi}_0}{\lam_r} + \sum_{n=1}^{N}\left(\lam_{\theta n}^e \pdv{\bar{\psi}_n}{\lam_{\theta n}^e} -  \lam_{rn}^e \pdv{\bar{\psi}_n}{\lam_{rn}^e}\right)\\
    &= \frac{\lam}{2}\tilde{\psi}'(\lam) + \sum_{n=1}^{N} \frac{\lam_{\theta n}^e}{2}\tilde{\psi}_n'(\lam_{\theta n}^e) \label{eq:app_s}
    \end{align}
\end{subequations}
where 
$ \tilde{\psi}_0(\lambda) = \bar{\psi}_0 \left(\lambda^{-2}, \lambda, \lambda\right)$ and $ \tilde{\psi}_n(\lam_{\theta n}^e) = \bar{\psi}_n \left({\lam_{\theta n}^e}^{-2}, \lam_{\theta n}^e, \lam_{\theta n}^e\right)$. Using \eqref{eq:app_s} and recalling that $\lam_{\theta n}^e = \lam / \lam_n^v$ from \eqref{eq_lambda_relation}, we arrive at \cref{eq:s_psi}. For the spherically symmetric deformation, \eqref{eq_app6} can be used to write
\begin{equation}
\label{eq_app98}
    \dot{\overline{{\lam_n^v}^2}} = 2 \lam_n^v \dot{\lam}_n^v = \frac{{\lam_n^v}^2}{\eta_{K_n}}\left(M_{\theta n} -\frac{1}{3}\left(M_{r n}+M_{\theta n}+M_{\phi n}\right)\right) 
\end{equation}
where $M_{\theta n} = M_{\phi n} = \lam_{\theta n}^e \pdv{\bar{\psi}_n}{\lam_{\theta n}^e}$ and  $M_{r n} = \lam_{r n}^e \pdv{\bar{\psi}_n}{\lam_{r n}^e}$. Simplyfing \eqref{eq_app98}, we get
\begin{equation}
\label{eq_app99}
    \dot{\lam}_n^v = \frac{{\lam_n^v}}{6 \eta_{K_n}}\left( \lam_{\theta n}^e \pdv{\bar{\psi}_n}{\lam_{\theta n}^e} - \lam_{r n}^e \pdv{\bar{\psi}_n}{\lam_{r n}^e}\right) = \frac{\lam}{12 \eta_{K_n}} \tilde{\psi}_n'(\lam_{\theta n}^e)
\end{equation}
Using \eqref{eq_app00} and the relation $\lam_{\theta n}^e = \lam / \lam_n^v$ in \eqref{eq_app99} gives us \eqref{Kinetic_Law_spec_gen}. Substituting \eqref{eq:psi12} in \eqref{eq_app99} and using the definition of $\tau_n$ in \eqref{general_evoln_viscous_cauchy}, we arrive at \cref{Kinetic_Law_spec}.

\subsection{Numerical integration}
For the numerical integration,  we choose to work with a fixed set of material points. The Lagrangian coordinate space $R\in[A\ B]$ is nonuniformly discretized such that the density of material points is higher closer to the cavity wall. The circumferential stretch for any material point is given by
\begin{equation}
\label{app:lam}
    \lam(R,t) = \left(1 + \left(\frac{A}{R}\right)^3(\lam_a^3(t)-1)\right)^{1/3}
\end{equation}
Given a prescribed cavity wall deformation $\lam_a(t)$, we time integrate the evolution equation for the viscous stretch $\lam_n^v(R,t)$, written in terms of $\lam(R,t)$,
\begin{equation}
        \dot{\lam}_{n}^v (R,t)  = \frac{4}{3 \tau_n}\frac{1}{\left({2\lam_n^v}^2 + {\lam_n^v}^{-4}\right)}\left( \frac{\lam^6 - {\lam_n^v}^6}{{\lam_n^v} \lam^4} \right) 
\end{equation}
The time integration is carried out using Matlab's `ode45' method and the tolerance and initial step values are chosen so as to ensure convergence for even small values of the parameters $\tau_n$. Separate time integrations are done for the expansion-relaxation and retraction-recovery part of every loading cycle using the initial conditions described in \Cref{subsec:in_conds}. Using \cref{eq:lam_1,eq:boundarystretches} and the integrated values for $\lam_n^v(R,t)$, the integration in \eqref{eq:Pfinal2} is carried out using a trapezoidal rule to calculate the gauge pressure $P(t)$. For evaluation of the inertial pressure term $P_{in}(t)$, we take $\rho=965$ kg/m${}^3$ and $\ddot{\lam}_a = 0$, but as noted earlier in \Cref{subsec:radialeqn}, $P_{in}(t)$ is insignificant compared to $S(t)$.

\subsection{Parameter fitting}
For every cycle, the equilibrated pressures at the start and end of the expansion-relaxation process, $P_{s_1}$ and $P_{s_2}$, are used to fit for the parameters $E$ and $A$ as described in \Cref{subsec:quasiparam}. To fit for the parameters $\alpha_1,\alpha_2, \tau_1$ and $\tau_2$ we make use of Matlab's nonlinear least squares method `$\textrm{lsqnonlin}$' using the `trust-region-reflective' algorithm. For the residual function, the pressure obtained from the numerical integration discussed in the previous section is subtracted from the experimental pressure data. First, control pressure profiles are generated by integrating the material model for given values of the viscoelastic parameters and it is ensured that the fitting method is able to exactly capture the chosen parameters. During this control testing, the convergence speed of the algorithm is optimised by experimenting with the scaling of the parameters using the `TypicalX' argument of `lsqnonlin'. The optimised fitting method is then used to fit the experimental data. It is also ensured that the fitting is roughly insensitive to the initial guess values for the parameters. The parameters are fitted for the expansion-relaxation part of the pressure profiles but the fitted parameters also predict the retraction-recovery profiles well. Since the expansion part of the loading cycle happens over a small fraction of the total time for expansion-relaxation, the pressure data is sampled non-uniformly for fitting so as to have a higher density of data points in the expansion part of the pressure-time curves. The fitting errors reported are, however, area integrals that are not affected by the re-sampling for fitting purposes. Also note that while using a single set of non-equilibrium parameters ($\alpha_1,\alpha_2, \tau_1$, $\tau_2$) to predict the pressure response for all the loading cycles, the parameter values for $E$ and $A$ are always taken to be the values fitted to the individual cycles.

\newpage
\section{Fitted parameters for all cycles}
\label{app:params}

 \begin{table}[H]
  \centering \footnotesize
  \label{TableAppendix_params}
\begin{tabular}{ccccccccccc}
\hline 
Sample &Cycle & $A$ (\si{mm}) & $E$ (\si{kPa}) & $\alpha_1$ &$\tau_1$ (\si{s})  & $\alpha_2$  & $\tau_2 $ (\si{s})  & $\epsilon_{exp}(\%)$ & $\epsilon(\%)$ & $\epsilon_{max}(\%)$ \\ 
\hline 
S50-00 & 1 & 0.353 & 14.71 & 0.597 & 2.76 & 0.097 & 154.63 & 0.23 & 0.18 & 1.45 \\ 
S50-00 & 2 & 0.356 & 14.77 & 0.789 & 1.54 & 0.109 & 129.07 & 0.37 & 0.27 & 2.17 \\ 
S50-00 & 3 & 0.358 & 14.77 & 0.927 & 1.00 & 0.116 & 123.18 & 0.47 & 0.36 & 2.67 \\ 
S50-00 & 4 & 0.360 & 14.79 & 1.246 & 0.56 & 0.128 & 110.56 & 0.50 & 0.44 & 4.06 \\ 
S50-00 & 5 & 0.363 & 14.83 & 1.599 & 0.32 & 0.140 & 97.16 & 0.63 & 0.53 & 5.87 \\ 
\hline 
S50-10 & 1 & 0.396 & 16.31 & 0.503 & 2.46 & 0.070 & 175.03 & 0.28 & 0.17 & 1.64 \\ 
S50-10 & 2 & 0.398 & 16.27 & 0.468 & 2.06 & 0.076 & 182.61 & 0.43 & 0.22 & 2.07 \\ 
S50-10 & 3 & 0.406 & 16.44 & 0.637 & 1.03 & 0.079 & 156.72 & 0.58 & 0.28 & 3.44 \\ 
S50-10 & 4 & 0.412 & 16.53 & 0.697 & 0.74 & 0.083 & 135.58 & 0.75 & 0.35 & 6.98 \\ 
S50-10 & 5 & 0.416 & 16.71 & 0.959 & 0.37 & 0.088 & 94.77 & 0.88 & 0.37 & 6.10 \\ 
\hline 
S50-20 & 1 & 0.402 & 20.43 & 0.309 & 2.44 & 0.040 & 148.00 & 0.46 & 0.11 & 1.56 \\ 
S50-20 & 2 & 0.403 & 20.46 & 0.315 & 1.65 & 0.044 & 139.70 & 0.52 & 0.14 & 1.71 \\ 
S50-20 & 3 & 0.404 & 20.48 & 0.404 & 0.85 & 0.051 & 125.91 & 0.79 & 0.18 & 3.17 \\ 
S50-20 & 4 & 0.406 & 20.54 & 0.487 & 0.44 & 0.058 & 108.79 & 0.85 & 0.22 & 4.30 \\ 
S50-20 & 5 & 0.408 & 20.64 & 0.262 & 0.67 & 0.053 & 112.77 & 0.94 & 0.23 & 2.91 \\ 
\hline 
S48-00 & 1 & 0.378 & 18.81 & 0.706 & 2.48 & 0.089 & 160.14 & 0.31 & 0.20 & 1.68 \\ 
S48-00 & 2 & 0.382 & 18.91 & 0.899 & 1.39 & 0.102 & 135.28 & 0.40 & 0.29 & 2.16 \\ 
S48-00 & 3 & 0.385 & 18.98 & 1.367 & 0.65 & 0.118 & 115.38 & 0.57 & 0.40 & 3.35 \\ 
S48-00 & 4 & 0.390 & 19.12 & 1.684 & 0.36 & 0.138 & 88.20 & 0.75 & 0.49 & 4.62 \\ 
S48-00 & 5 & 0.391 & 19.09 & 1.960 & 0.25 & 0.138 & 102.76 & 0.87 & 0.59 & 6.71 \\ 
\hline 
S48-10 & 1 & 0.348 & 17.89 & 0.135 & 9.34 & 0.052 & 179.11 & 0.37 & 0.08 & 1.38 \\ 
S48-10 & 2 & 0.351 & 17.98 & 0.203 & 4.34 & 0.057 & 139.47 & 0.50 & 0.10 & 1.93 \\ 
S48-10 & 3 & 0.349 & 17.84 & 0.273 & 2.56 & 0.066 & 169.41 & 0.66 & 0.16 & 2.56 \\ 
S48-10 & 4 & 0.355 & 17.98 & 0.380 & 1.27 & 0.070 & 131.88 & 0.83 & 0.21 & 3.19 \\ 
S48-10 & 5 & 0.357 & 18.03 & 0.644 & 0.49 & 0.077 & 112.00 & 0.81 & 0.29 & 3.04 \\ 
\hline 
S48-20 & 1 & 0.420 & 24.13 & 0.376 & 2.32 & 0.033 & 200.36 & 0.36 & 0.10 & 1.33 \\ 
S48-20 & 2 & 0.424 & 24.22 & 0.440 & 1.47 & 0.039 & 178.49 & 0.55 & 0.13 & 2.32 \\ 
S48-20 & 3 & 0.428 & 24.33 & 0.505 & 0.94 & 0.041 & 172.33 & 0.69 & 0.18 & 2.52 \\ 
S48-20 & 4 & 0.433 & 24.47 & 0.517 & 0.74 & 0.039 & 186.29 & 0.81 & 0.23 & 2.77 \\ 
S48-20 & 5 & 0.438 & 24.63 & 0.621 & 0.43 & 0.046 & 146.80 & 0.81 & 0.28 & 3.73 \\ 
\hline 
\end{tabular}
 \end{table}


\end{document}